\documentclass[aps,pre,twocolumn,groupedaddress,showpacs]{revtex4}
\usepackage{graphicx}
\usepackage{latexsym,lscape,revsymb,natbib}
\usepackage{dcolumn}
\usepackage{bm}

\newcommand{\be}{\begin{eqnarray}}
\newcommand{\ee}{\end{eqnarray}}
\newcommand{\bi}{\bibitem}

\newcommand{\benl}{\begin{eqnarray*}}
\newcommand{\eenl}{\end{eqnarray*}}
\begin{document}

\title{Equilibrium structures and flows of polar and nonpolar liquids and their mixtures in carbon nanotubes with rectangular cross sections}

\author{L.V. Mirantsev}
\email[author to whom correspondence should be addressed. Email address:]{miran@mail.ru}
\homepage[]{www.ipme.ru}
\author{A. K. Abramyan} 	
\affiliation{Institute for Problems of Mechanical Engineering, Russian Academy of Sciences,
 199178, Bolshoi 61, V. O., St. Petersburg, Russia}

\date{\today}


\begin{abstract}

Molecular dynamics (MD) simulations of  equilibrium structures and flows of polar water, nonpolar argon and methane, and mixtures of water and methane
confined by single - walled carbon nanotubes (SWCNTs) with different rectangular cross sections have been performed.
The results of these simulations show that equilibrium structures and flows of all confined liquids significantly depend not only on the shape of 
the SWCNT's rectangular cross sections but also on the types of liquids inside SWCNTs.

\end{abstract}
\maketitle
\section{Introduction}\label{sec:introduction}

During the last 20 - 30 years, carbon nanotubes (CNTs) attract a huge interest of the international research community.
There are two types of CNTs, namely, single - walled  carbon nanotubes (SWCNTs) and multi - walled carbon nanotubes (MWCNTs) which are structures
composed of several concentric SWCNTs with different chiral angles \cite{01,02,03,04,05}.  Diameters of CNTs range from a few to tens nanometers, and their
lengths may reach several microns\cite{06,07,08,09}. Both SWCNTs and MWCNTs exhibit superior mechanical, electronic, thermal and transport properties  
\cite{10,11,12,13,14,15,16,17,18,19,20,21,22}. For example, experimental investigations and computer simulations revealed that a pressure drop 
driven fluid flows through CNTs could be much faster relative to predictions of classic hydrodynamics. 
Due to their unique properties, CNTs are widely used in many electronic, medical, space, and military applications \cite{23,24,25}. Furthermore, 
membranes and films formed dy CNTs can be also used for a water purification and separation of various organic contaminants from the water \cite{26,27,28}. 

The most studies of CNTs have been dedicated to controlling their length \cite{29,30}, wall number \cite{31,32}, diameter
 \cite{33,34,35,36,37,38}, and chirality \cite{39,40}, and in all these studies CNTs had circular cross sections. However, in \cite{41}, 
a possibility of formation of CNTs with rectangular cross sections was discussed, and, recently, it was reported in \cite{42} that such carbon nanotubes 
can be really formed. In our previous paper \cite{43}, using molecular dynamics (MD) simulations, we performed study of equilibrium behaviors and flows
of polar and nonpolar liquids confined by SWCNTs with square cross section and showed that equilibrium fluid structures inside such SWCNTs and fluid flows 
through them can be strongly different from those in SWCNTs with circular cross sections. Nevertheless, rectangular cross sections of CNTs could have
shapes different from the square one, and equilibrium and dynamic behaviors of various fluids inside such CNTs could be significantly different from those 
in CNTs with square cross sections. 

In the present paper, using MD simulations, we study equilibrium structures and flows of polar and nonpolar liquids and their mixtures confined 
by SWCNTs having rectangular cross sections with different ratios between their sides, namely, 1 : 1 (square cross section), 1 : 2, and 1 : 4. All 
these nanotubes have the same length and the cross section area. In our simulations, the polar fluid is the water, and the nonpolar ones are methane and
argon in their liquid phase. It has been found that equilibrium structures of above mentioned confined liquids depend strongly on the shapes  of cross 
sections of SWCNTs. In addition, these equilibrium structures and flows of such fluids are very sensitive to a polarity of the fluid and interactions
 between their atoms (molecules) and interactions between liquid particles and boundary wall carbon atoms. 

\section{Simulation details}\label{sec:simulation details}

As said above, using MD simulations, we investigate  static and dynamic behavior of polar and nonpolar model fluids and their mixtures confined 
by carbon nanotubes with rectangular cross sections. As in \cite{43,44}, we use a very simple model of polar fluid in which molecules are assumed to be point-like 
particles possessing a permanent dipole moment. These particles interact with each other via the short-range Lennard-Jones (LJ) pairwise potential
   
\begin{equation}\label{eq:LM1}
U_{LJ}(r_{ij}) = 4\epsilon_{ij}\lceil\left(\sigma_{ij}/r_{ij}\right)^{12} - \left(\sigma_{ij}/r_{ij}\right)^{6}\rceil,
\end{equation}

where $\epsilon_{ij}$ and $\sigma_{ij}$ are the strength and characteristic length, respectively, for the LJ interaction between $i$-th and $j$-th 
molecules, and $r_{ij}$ is the distance between them , and the dipole-dipole interaction potential

\begin{equation}\label{eq:LM2}
U_{dd}(r_{ij}) = (\vec d_i\cdot\vec d_j)/r^3_{ij} - 3(\vec d_i\cdot\vec r_{ij})(\vec d_j\cdot\vec r_{ij})/r^5_{ij},
\end{equation}

where $\vec d_{i}$ is the dipole moment of the $i$-th molecule. Since the polar molecules are considered as the water ones, the LJ interaction 
constants $\epsilon_{ij}$ and  $\sigma_{ij}$ in Eq. (1) should be similar to those for the LJ interactions between oxygen atoms of the $i$-th and $j$-th 
water molecules. These constants, $\epsilon_{H2O}$ and $\sigma_{H2O}$, are taken from the well known model for water molecules \cite{45,46}, 
and they are equal to $\epsilon_{H2O} = 1.083\times 10^{-14}$ erg and $\sigma_{H2O} = 3.166${\AA},respectively. As in \cite{43,44}, 
an effective dipole moment of the water molecule is set $d_w = 1.89\times 10^{-18}$ g$^{1/2}$cm$^{5/2}$s$^{-1}$.
An interaction between water molecules and carbon atoms of SWCNTs is described by the LJ potential similar to Eq.(1)in which the 
interaction constants $\epsilon_{H2O}$ and $\sigma_{H2O}$ are replaced by $\epsilon_{CH2O}$ and $\sigma_{CH2O}$, respectively. 
All these constants are taken from \cite{45}. As for nonpolar argon and methane molecules, they are described in framework 
of the united atom model of methane \cite{17} and the well known model of argon atoms \cite{47}. According to this models, the argon atoms $Ar$ and the 
methane molecules $CH4$ also interact with each other and with carbon atoms of SWCNTs via the LJ potential (1). The corresponding interaction constants
for methane $\epsilon_{CH4}$, $\sigma_{CH4}$, $\epsilon_{CCH4}$, and $\sigma_{CCH4}$ are taken from \cite{17}, and the constants $\epsilon_{Ar}$, 
$\sigma_{Ar}$ for argon atoms are taken from \cite{47}. As for interactions between argon and carbons atoms, the corresponding constants   
$\epsilon_{CAr}$, and $\sigma_{CAr}$ are determined by means of the Lorentz - Berthelot rules. 
The total force $\vec F_{i}$ and the total torque $\vec\tau_{i}$ acting on the i-th fluid molecule due to interactions with other molecules,
and equations of motion of this molecule are given by Eqs. (3) - (8) in \cite{44}. All simulations are performed in the $NVT$ ensembles, and,  
at each time step, the equations of motion of fluid molecules are solved numerically by the standard method described in \cite{48}. The temperatures 
of the systems under consideration are kept constant ($T = 300 K$ for water and its mixture with methane, $T = 108 K$ for liquid phase of methane, and 
$T = 85 K$ for liquid phase of argon) by employment of the Berendsen thermostat \cite{49}.

\section{Simulation results and discussion}\label{sec:results of simulations and discussion}

First of all, we studied equilibrium static structures of the polar water molecules, nonpolar argon atoms and methane molecules, and the mixture of water 
and methane molecules in SWCNTs mentioned in 
Introduction. The lateral projections and cross sections of these SWCNTs are shown in figures 1 (a) - 1(c). 

\begin{figure}[t]\centering
\includegraphics[width = 8cm]{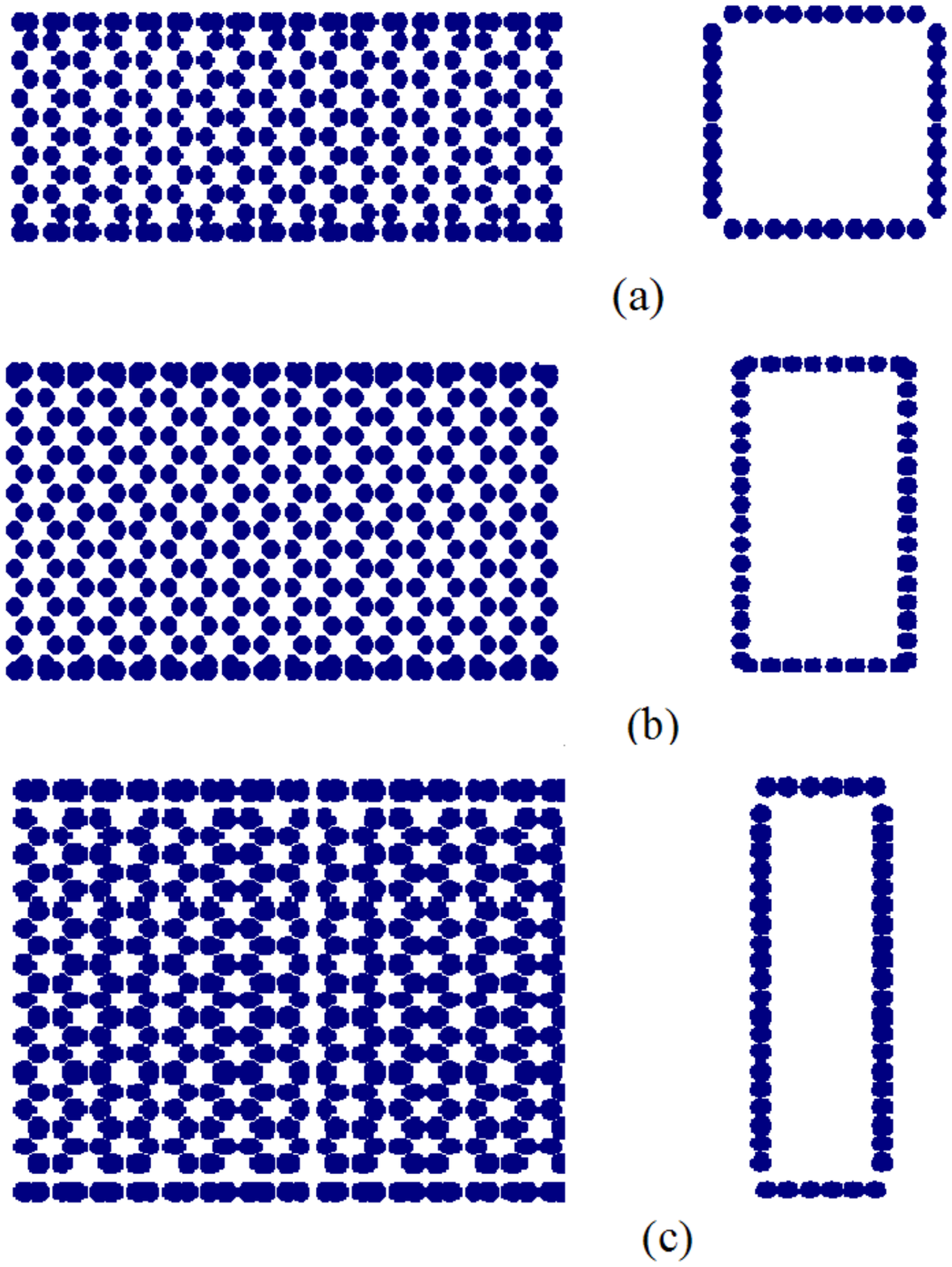}\kern-24mm
\caption{\label{f01} The lateral projections and cross sections of SWCNTs used in our MD simulations. a - SWCNT with square cross section; b - SWCNT with 
rectangular cross section and with the ratio between its sides 1 : 2; c - SWCNT with rectangular cross section and with the ratio between its sides 1 : 4.}

\end{figure}

All these carbon nanotubes have the same length $L = 3.8 nm$ and the cross section area $S = 1.77 nm^2$, and their bounding walls have the graphene 
crystalline structure. Since the carbon atoms in CNTs are connected to 
each other with very strong covalent bonds \cite{50} with the interaction constants much larger than the LJ interaction constants 
$\epsilon_{CH2O}$,$\epsilon_{CAr}$ and $\epsilon_{CCH4}$, these atoms are considered to be fixed at their equilibrium sites. This approximation is supported
by estimations of thermal vibrations of SWCNTs performed in \cite{51}. According to these estimations, average amplitudes of such vibrations are of the 
order of $\sim 0.01$ nm that is significantly smaller than typical molecular sizes. In order to obtain equilibrium static structures 
of the water, argon, methane, and the mixture of the water and methane molecules  inside the above mentioned SWCNTs, we performed MD simulations of free 
(without an external pressure drop) permeations of water and methane molecules, and argon atoms into these SWCNTs. These simulations started from the 
initial configuration schematically depicted in 
figure 2(a). This configuration is made as follows: initially we have two reservoirs of water or methane molecules, and argon atoms separated from each 
other by the wall consisted of carbon atoms. Then, we make a channel in this wall and insert the corresponding SWCNT into the channel. Now, this SWCNT 
connects two reservoirs with each other and liquid atoms (molecules) can freely permeate into the SWCNT without any external action. The same 
procedure is performed for the mixture of the water and methane molecules. In this case, both reservoirs contain initially equal numbers of the water and 
methane molecules. After running during about 100000 time steps (one time step was equal to 0.001 ps), one can reach the equilibrium configuration 
schematically shown in figure 2(b). 

\begin{figure}[t]\centering\vspace{-10mm}
\includegraphics[width = 8cm,clip=on]{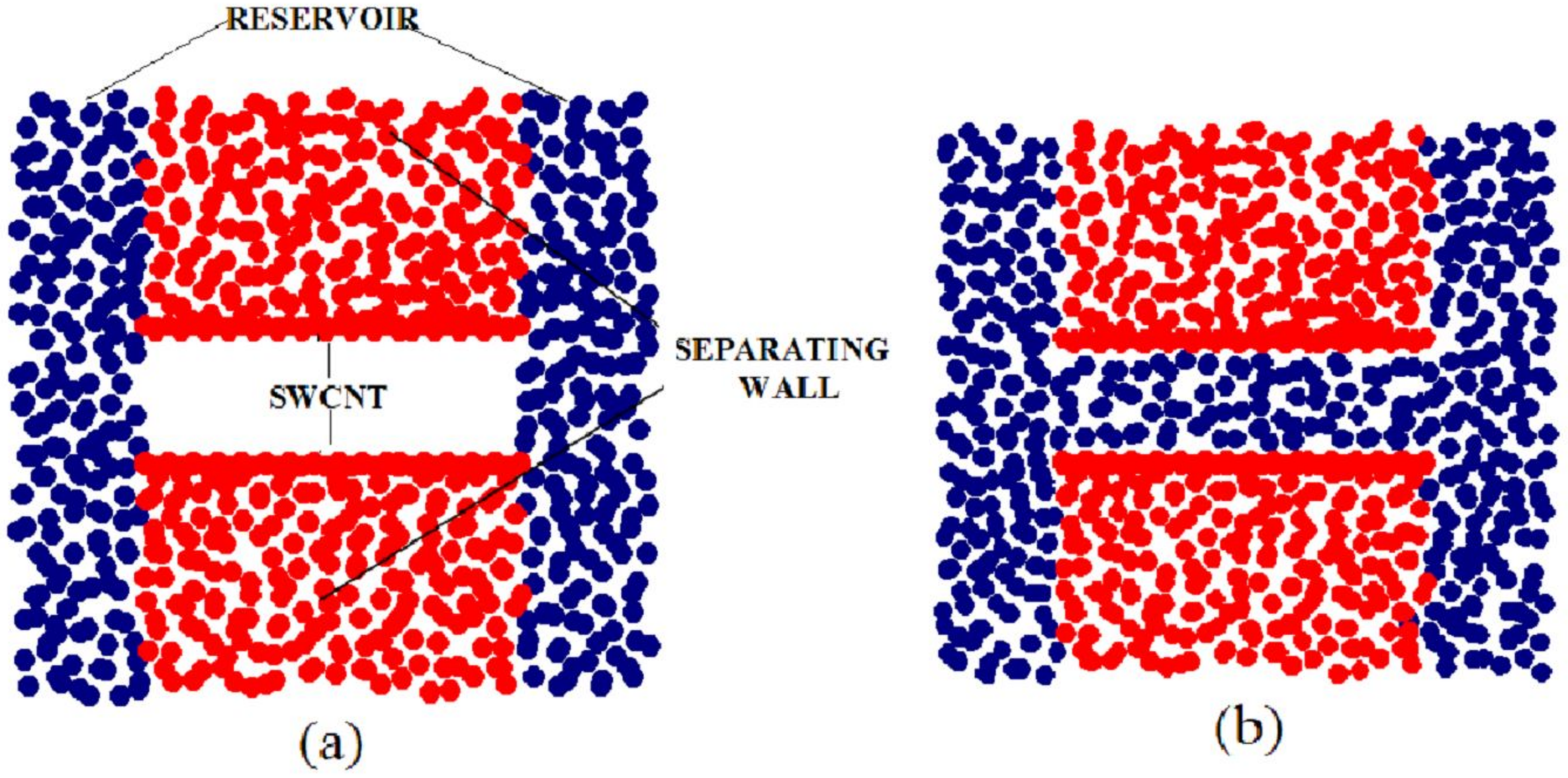}\vspace{-11mm}
\caption{\label{f02} The lateral projection of initial and final equilibrium configurations considered in our MD simulations. a - initial configuration;
b - final equilibrium configuration. Blue circles are liquid atoms (molecules), and red ones are carbon atoms.}\end{figure}

During simulation processes, the system under consideration is placed 
within the parallelepiped simulation box of $6.332\times 6.332\times 6.332 nm^3$ in size, and the periodic boundary conditions 
\cite{48} were imposed on the system in $x$ $y$, and $z$ directions.

Figures 3a - 3l demonstrate cross sections of equilibrium configurations of water molecules (figures 3a - 3c), argon atoms (figures 3d - 3f), methane
molecules (figures 3g - 3i), and mixture of water and methane molecules (figures 3j - 3l) inside SWCNTs with the described above rectangular cross sections. 
The corresponding density profiles along $z$ (vertical) and $y$ (lateral) directions perpendicular to the tube $x$ axes are shown in figures 4 - 7. If we
look at figures 3a - 3l, it is clearly seen that, for all types of liquid atoms (molecules), the main features of the corresponding equilibrium structures
are similar, and these features are defined by the shapes of SWCNT cross sections. For example, all liquid atoms (molecules) inside SWCNTs with square 
cross sections form structures with square - like cross sections which are reduced replicas of the SWCNTs ones (see figures 3a, 3d, 3g, 3j). All liquid 
atoms (molecules) inside SWCNTs with rectangular cross stctions having the ratio 1 : 2 between their sides form equilibrium structures consisting of two
planes parallel to the vertical bounding walls (see figures 3b, 3e, 3h, 3k). Finally, all liquid atoms (molecules)inside SWCNTs with rectangular cross
sections having the ratios 1 : 4 between their sides form equilibrium 2D structures in a form of the plane parallel to the vertical bounding walls (see
figures 3c, 3f, 3i, 3l). Thus, the shape of the rectangular SWCNT cross sections plays a dominant role in a formation of equilibrium structures of liquid
atoms (molecules) inside SWCNTs.

However, it is simultaneously seen that the equilibrium structures depicted in figures 3a - 3l demonstrate certain additional features depending on 
characteristics of concrete liquid atoms (molecules) under consideration. It is easily seen that, inside all SWCNTs, liquid structures formed by argon 
atoms are most ordered. For example, argon atoms inside SWCNT with square cross section form 9 well ordered chains parallel to the tube axis 
(see figure 3d). One of these chains
coincides with the tube axis, and other 8 chains are disposed on bounding surfaces of imaginable parallelepiped inside this SWCNT. Simultaneously, 
methane molecules inside the same SWCNT form similar structure but with more smeared chains (see figure 3g). This fact is also 
reflected in figures 5a and 6a, which demonstrate the density profiles along $z$ axis for argon atoms and methane molecules inside the same SWCNT with 
square cross section. One can see that the density profile for argon atoms exhibits three peaks having almost the same height and width equal to about
0.6 $\sigma_{Ar}$. The analogous profile for methane molecules has also three peaks, but the central one is noticeably lower than two others, and the width
of these peaks is about 0.6 $\sigma_{CH4}$ that is about 10 percent larger than that of the analogous peaks in the density profile for argon atoms. If 
we look at figure 3a which exhibits the cross section of the equilibrium water structure inside the SWCNT with the square cross section and at figure 4a
demonstrating the corresponding density profile along $z$ axis, then we find that this structure is much more disordered relative to those depicted in 
figures 3d and 3g for argon atoms and methane molecules, respectively. Perhaps, it is due to the Coulomb - like dipole - dipole interactions between polar
water molecules which do not occur in ensembles of nonpolar argon atoms and methane molecules. According to the Earnshaw theorem \cite{52}, an ensemble of 
particles interacting via Coulomb - like forces cannot be maintained in a stable stationary equilibrium configurations. Thus, the well ordered structures of 
water molecules inside SWCNTs cannot exist for a sufficiently long times. If we look at figure 3j, which demonstrates the cross section of the equilibrium 
structure formed by the mixture of water and methane molecules inside SWCNT with square cross section, we can find that this structure resembles the above
mentioned structure formed by the pure methane inside the same SWCNT. This fact is also confirmed by the density profiles for methane and water molecules 
depicted in figure 7a. The density profile for methane molecules (curve 2) exhibits three well developed peaks of almost similar height, whereas the 
analogous profile for water molecules (curve 1) demonstrates two well developed peaks at the edges and the strongly smeared central one. This fact can be 
explained by that the interaction constants $\epsilon_{CH4}$ and $\epsilon_{CCH4}$ for interactions between methane molecules and those between methane 
molecules and boundary wall carbon atoms, respectively, are significantly larger than analogous interaction constants $\epsilon_{H2O}$ and 
$\epsilon_{CH2O}$ for water molecules. So, if we look at
figures 3b, 3c, 3e, 3f, 3h, 3i, 3k, 3l, which demonstrate equilibrium structures formed by water molecules, argon atoms, methane molecules, and the mixture 
of water and methane molecules inside SWCNTs with rectangular cross sections having the ratios between their sides 1 : 2 and 1 : 4, and at figures 4b - 4e,
5b - 5e, 6b - 6e, 7b - 7e demonstrating the corresponding equilibrium density profiles, we can conclude that the said above for equilibrium structures 
inside SWCNT with square cross section is valid for equilibrium structures in all SWCNTs with rectangular cross sections. The most ordered structures are 
exhibited by argon atoms, whereas water molecules form most disordered structures, and the positional order of structures formed by methane molecules and
the mixtures $H2O + CH4$ have an intermediate positional order.

It should be also noted that the shape of the SWCNT's cross section and the interaction constants $\epsilon$ and $\sigma$ play very important role not only in 
equilibrium structures and average liquid densities inside SWCNTs but also in a compositions of the mixture of water and methane molecules inside different
SWCNTs. For example, inside SWCNT  with square cross section, the ratio between numbers of water and methane molecules is equal to 43 : 59, whereas inside
SWCNTs with rectangular cross sections these molecules occur in almost equal proportions, namely, 52:49 for SWCNT with ratio between its sides 1 : 2 and 
43 : 46 for SWCNT with the side ratio 1 : 4. This result can be qualitatively explained as follows. In the case of SWCNT with square cross section, the 
distance between bounding walls is large enough for both water and methane molecules. Therefore, the strengths of interactions between liquid molecules and
bounding walls, $\epsilon_{CH2O}$ and $\epsilon_{CCH4}$, play a main role in their penetration into SWCNTs, and the effective sizes of these molecules,         
$\sigma_{H2O}$ and $\sigma_{CH4}$, play a minor role. Since the interaction constant $\epsilon_{CCH4}$ for the interaction between methane molecules and
bounding wall carbon atoms is noticeably larger than the analogous interaction constant $\epsilon_{CH2O}$ for the water molecules, then the number of methane 
molecules penetrating into SWCNT with square cross section is larger than that of water molecules. For SWCNTs with rectangular  cross sections having
ratios between their sides equal to 1 : 2 and 1 : 4, respectively, the distances between lateral bounding walls along y direction become sufficiently small
for larger methane molecules and and not so small for smaller water molecules (effective diameter $\sigma_{CH4}$ of the methane molecule is about 10 percent
larger than the analogous diameter $\sigma_{H2O}$ of the water molecule). Thus, for these SWCNTs, a competition between interactions of different liquid 
molecules with bounding wall atoms and their molecular sizes occurs, and these competition equalizes concentrations of water and methane molecules inside 
such SWCNTs.
 
\begin{figure}[t]
\includegraphics[width = 8cm]{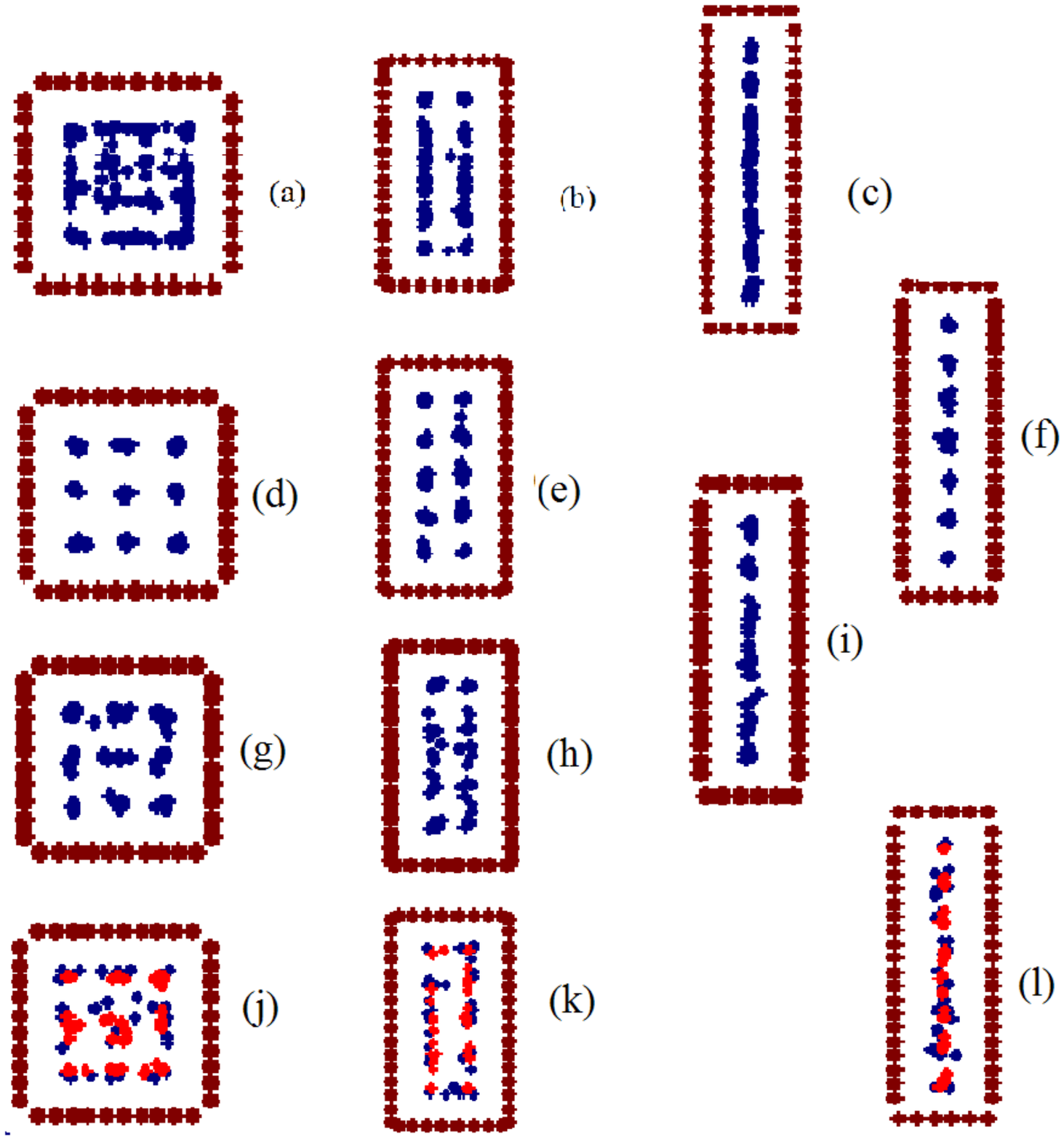}
\caption{\label{f03} Cross sections of equilibrium  structures of water molecules (3a - 3c), argon atoms (3d - 3f), methane molecules (3g - 3f),
and the mixture of the water and methane molecules (3j - 3l) inside SWCNTs with rectangular cross sections depicted in figures 1a - 1c. In all figures
brown circules denote the bounding wall carbon atoms. In figures 3a - 3i, blue circules denote liquid atoms (molecules). In figures 3j - 3l, blue and red 
circules denote the water and methane molecules, respectively. The equilibrium ratios between the water and methane molecules are 43 : 59, 52 : 49, 
and 43 : 46 for SWCNTs with rectangular cross sections with ratious between their sides equal to 1 : 1, 1 : 2, and 1 : 4, respectively.} 
\end{figure}

\begin{figure*}[t]\centering
\includegraphics[width = 8cm]{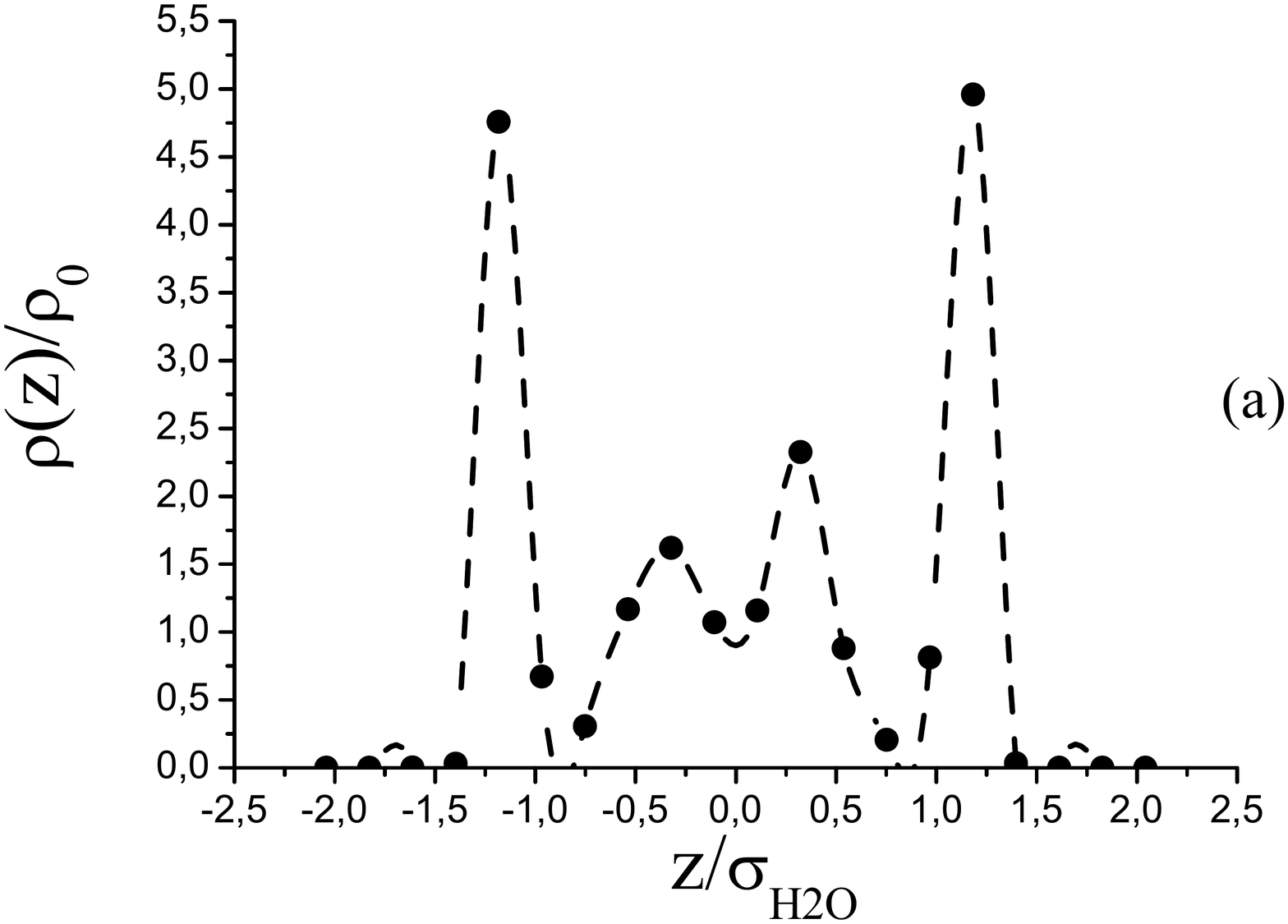}\kern6mm\includegraphics[width = 8cm]{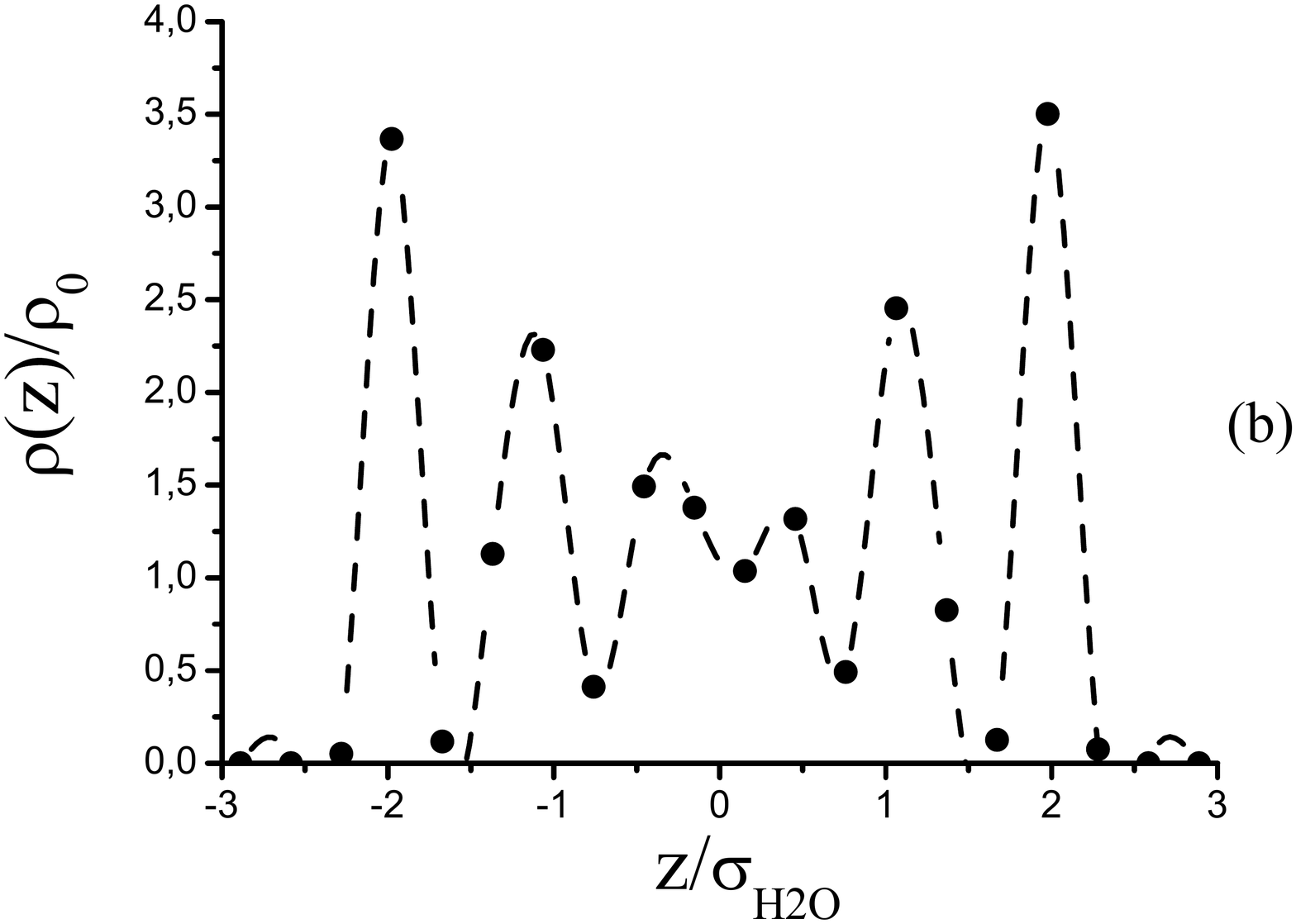}\\
\includegraphics[width = 8cm]{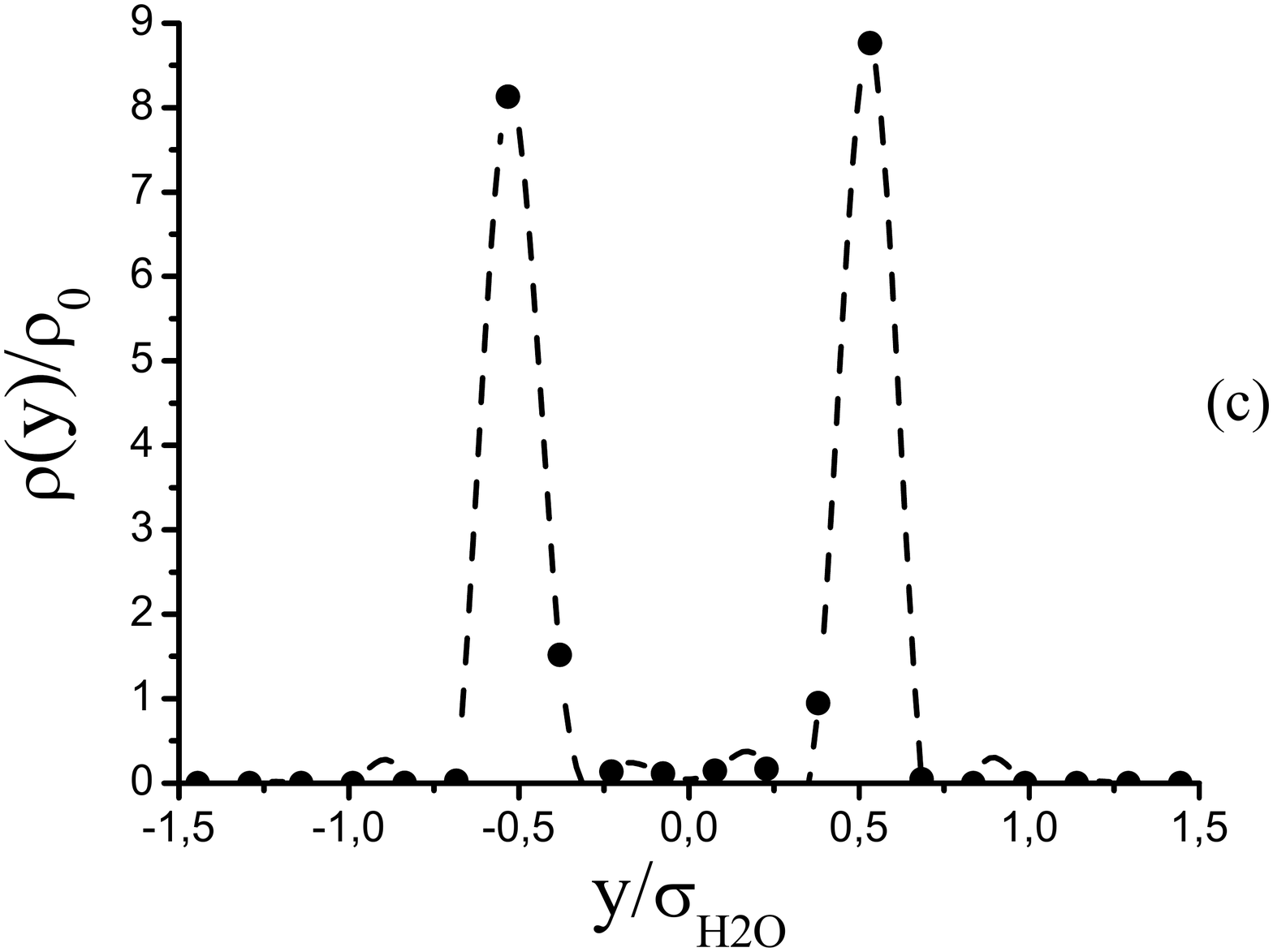}\kern6mm\includegraphics[width = 8cm]{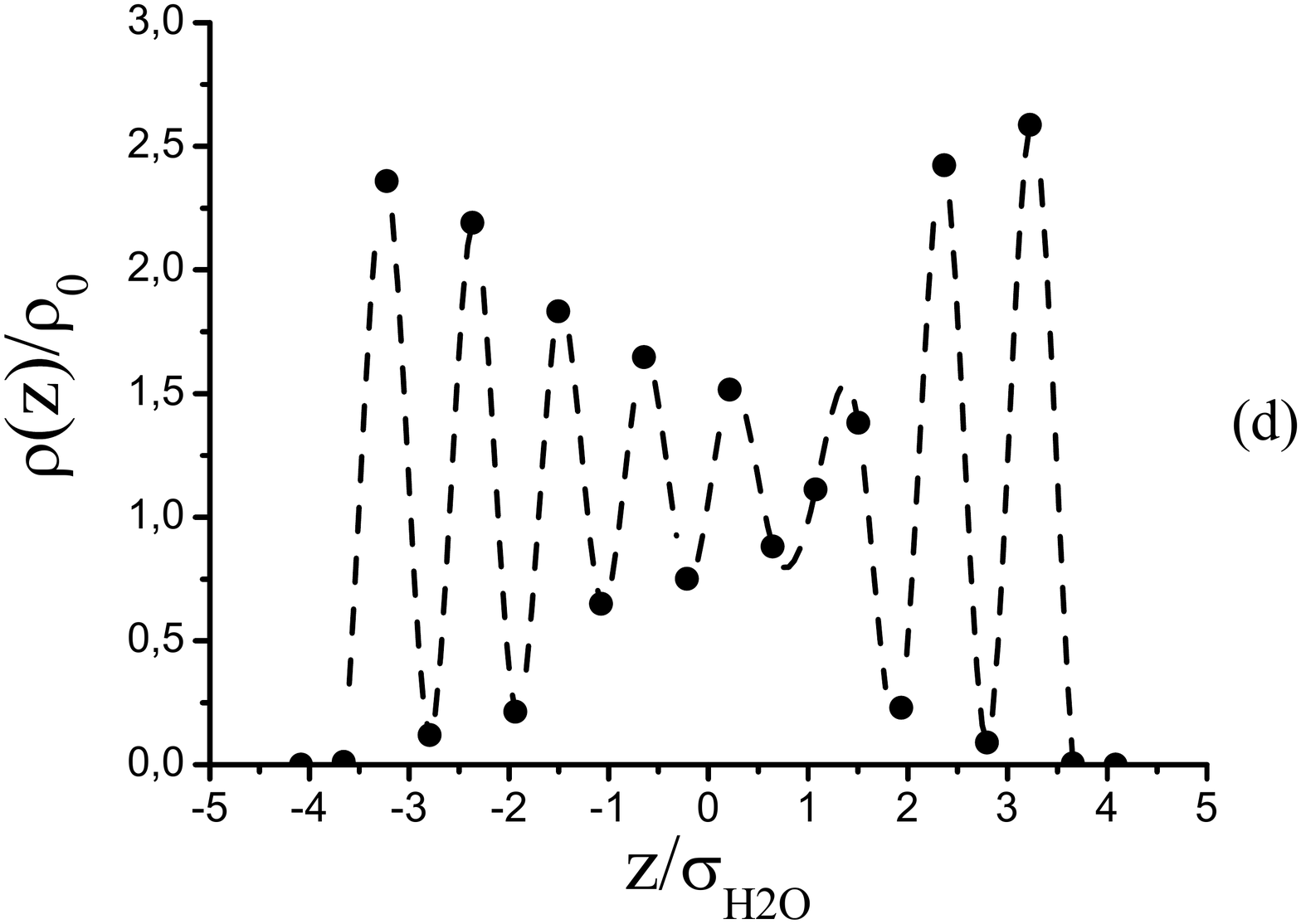}\\
\includegraphics[width = 8cm]{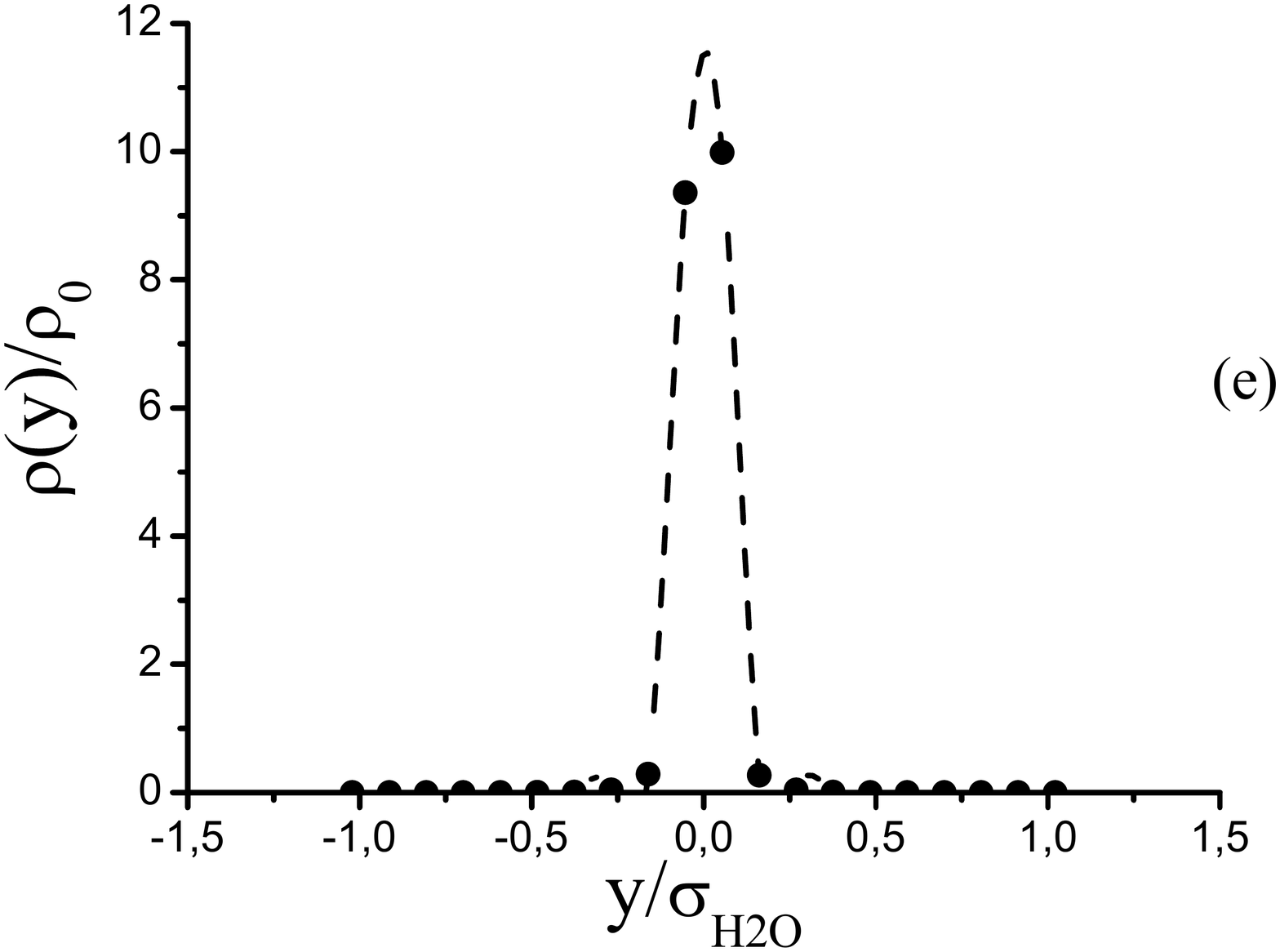}
\caption{\label{f04} Equilibrium density profiles for the water inside SWCNTs with different rectangular cross sections. a - The density profile along 
$z$ axis for SWCNT with square cross section. For symmetry reasons the analogous profile along $y$ axis should be similar. $\rho_0 = 0.61745$ is the 
average density in the tube in reduced MD units. b - The analogous profile for SWCNT with rectangular cross section having the ratio between its sides 
1 : 2. c - The density profile along $y$ axis for the same SWCNT. $\rho_0 = 0.5996$. d - The density profile along $z$ axis for SWCNT with rectangular ( cross section having the ratio between its sides 1 : 4. e - The density profile along $y$ axis for the same SWCNT. $\rho_0 = 0.44619$.}
\end{figure*}

\begin{figure*}[t]\centering
\includegraphics[width = 8cm]{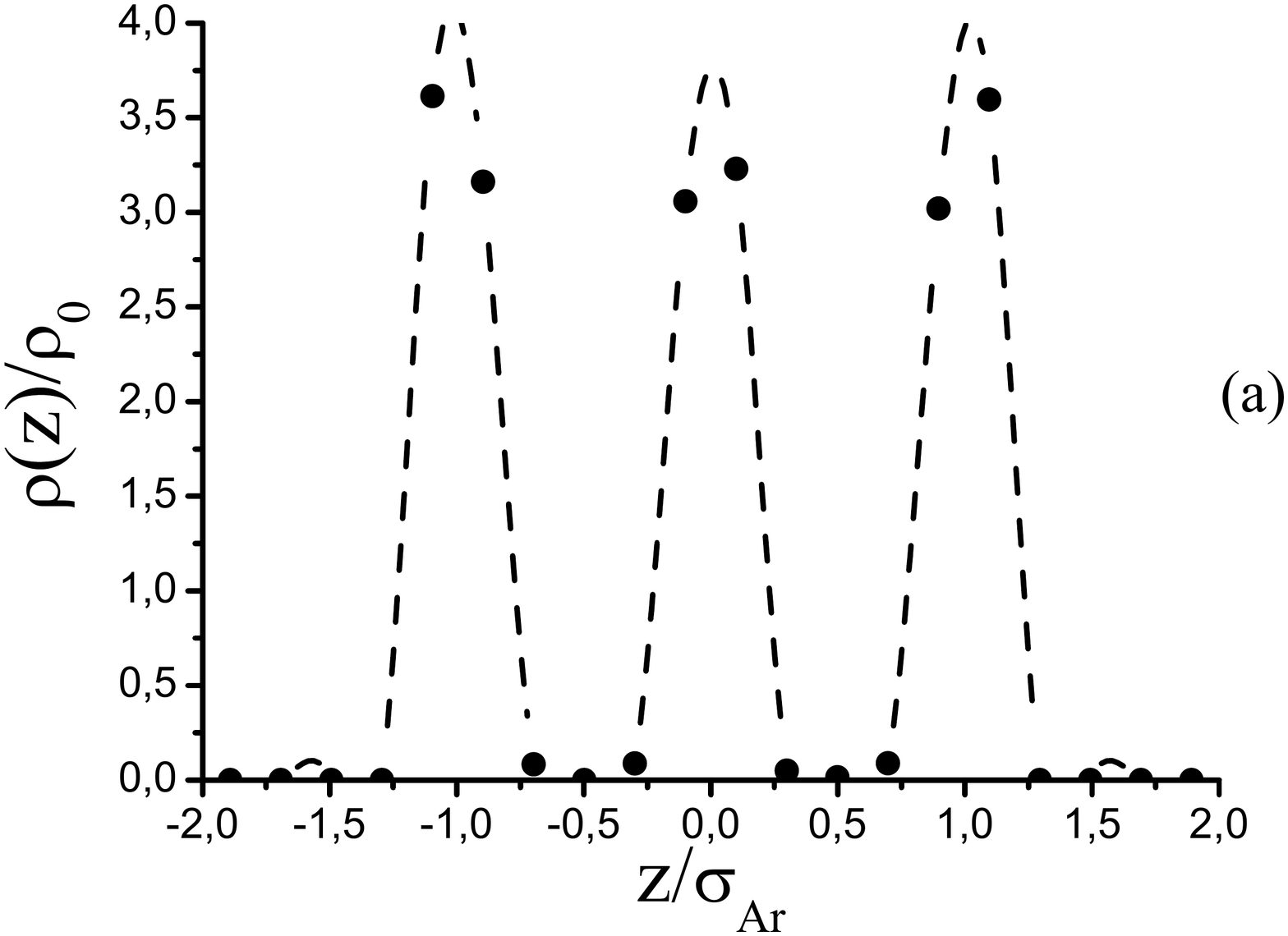}\kern6mm\includegraphics[width = 8cm]{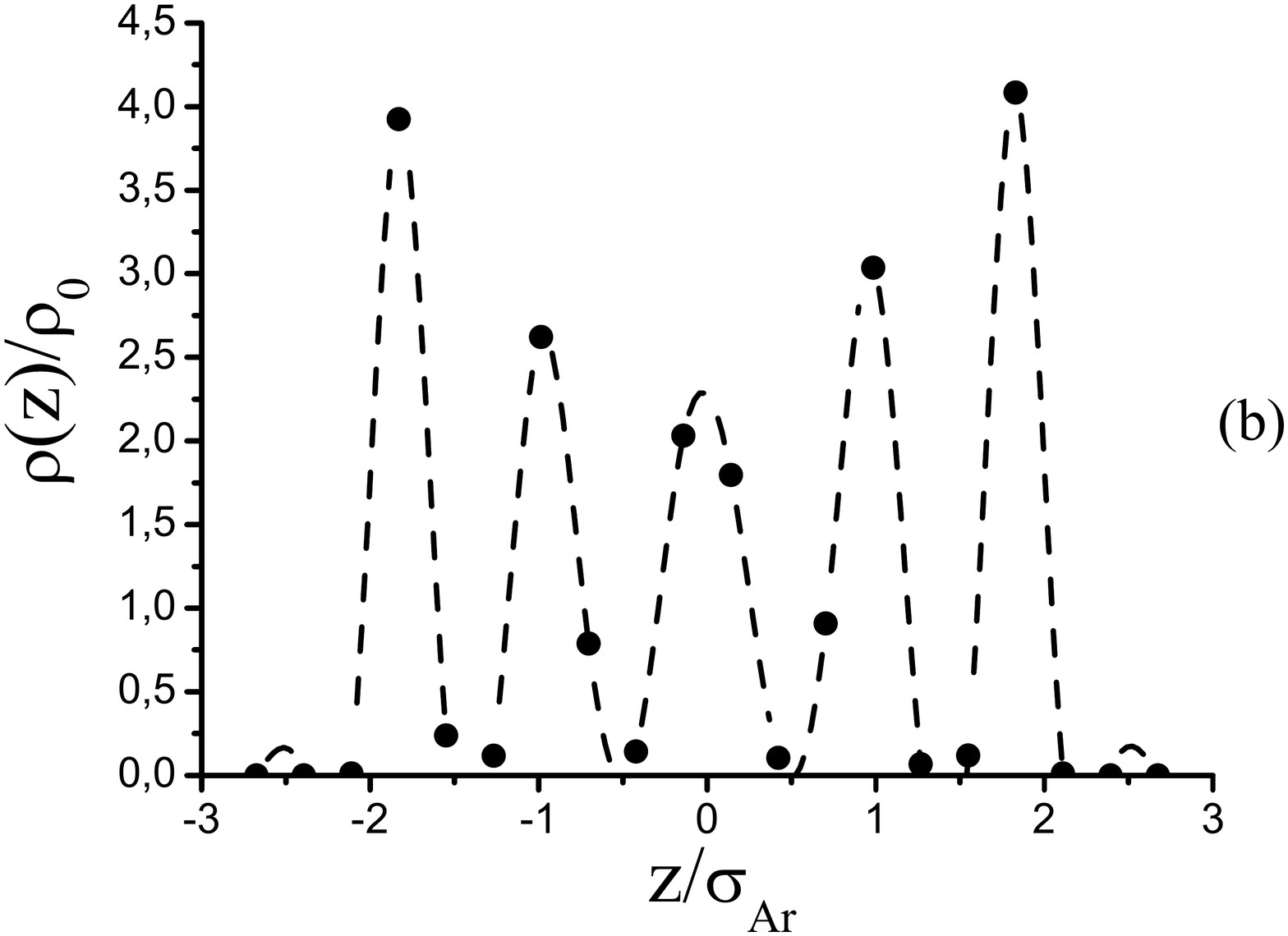}\\
\includegraphics[width = 8cm]{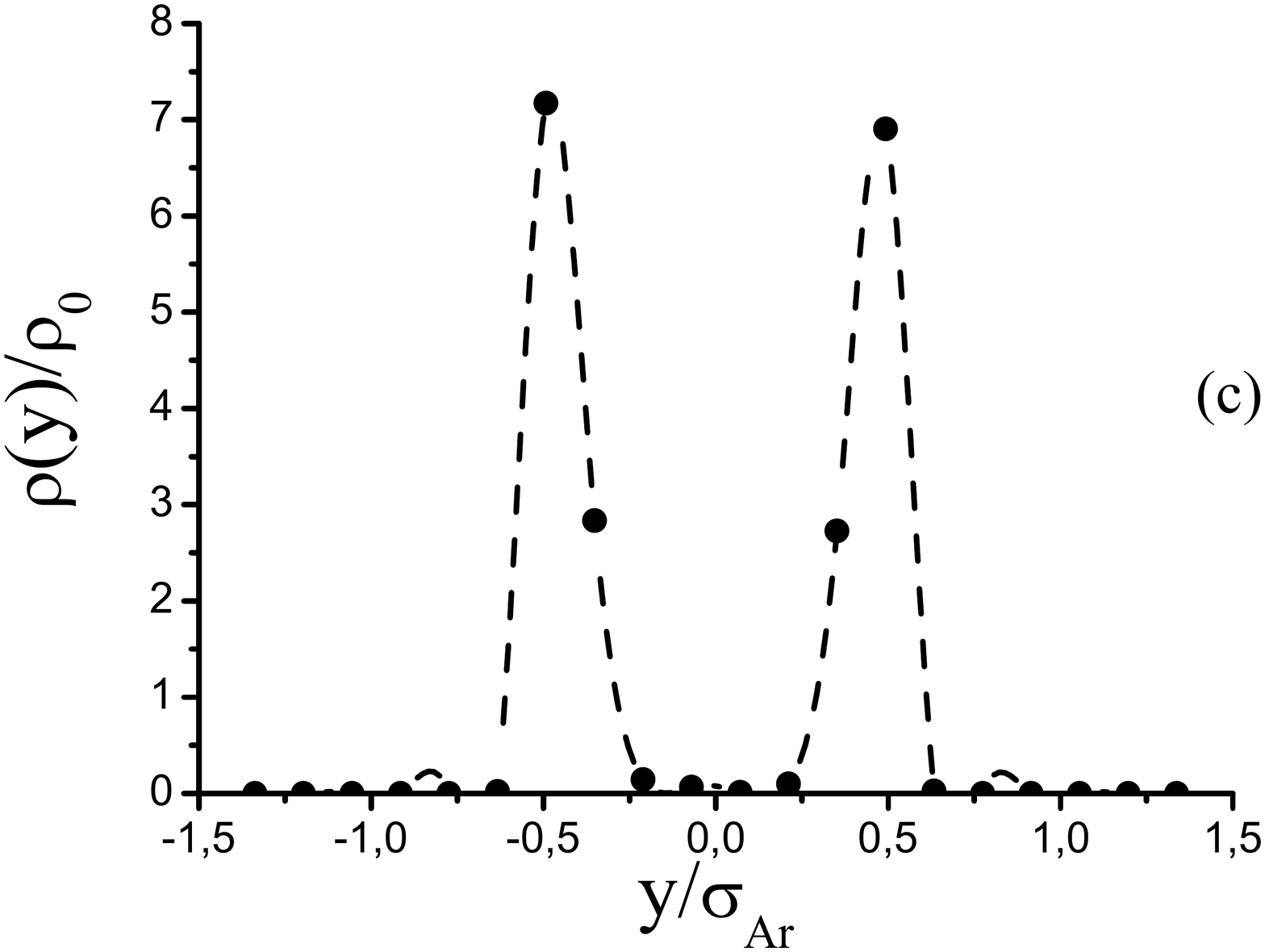}\kern6mm\includegraphics[width = 8cm]{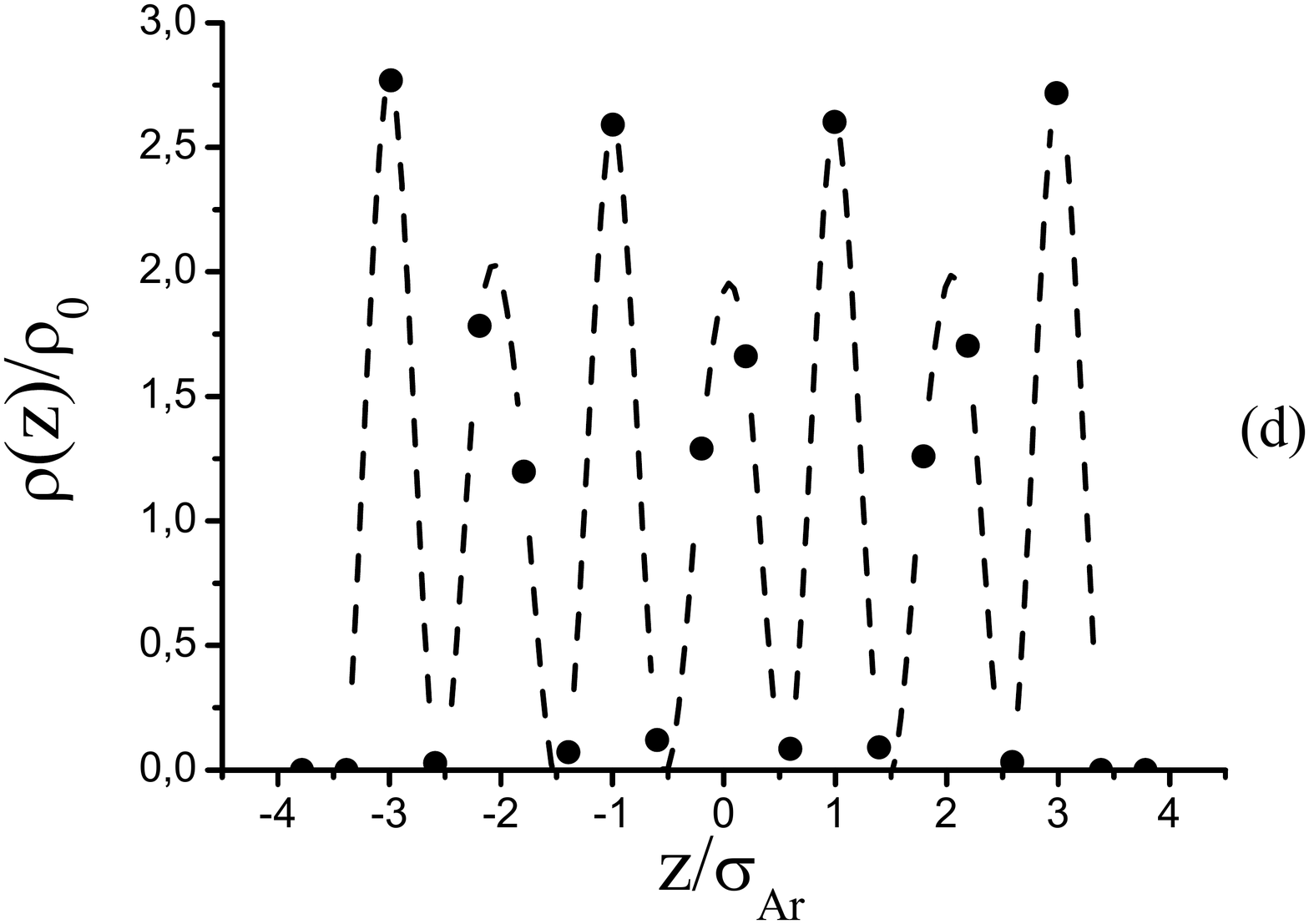}\\
\includegraphics[width = 8cm]{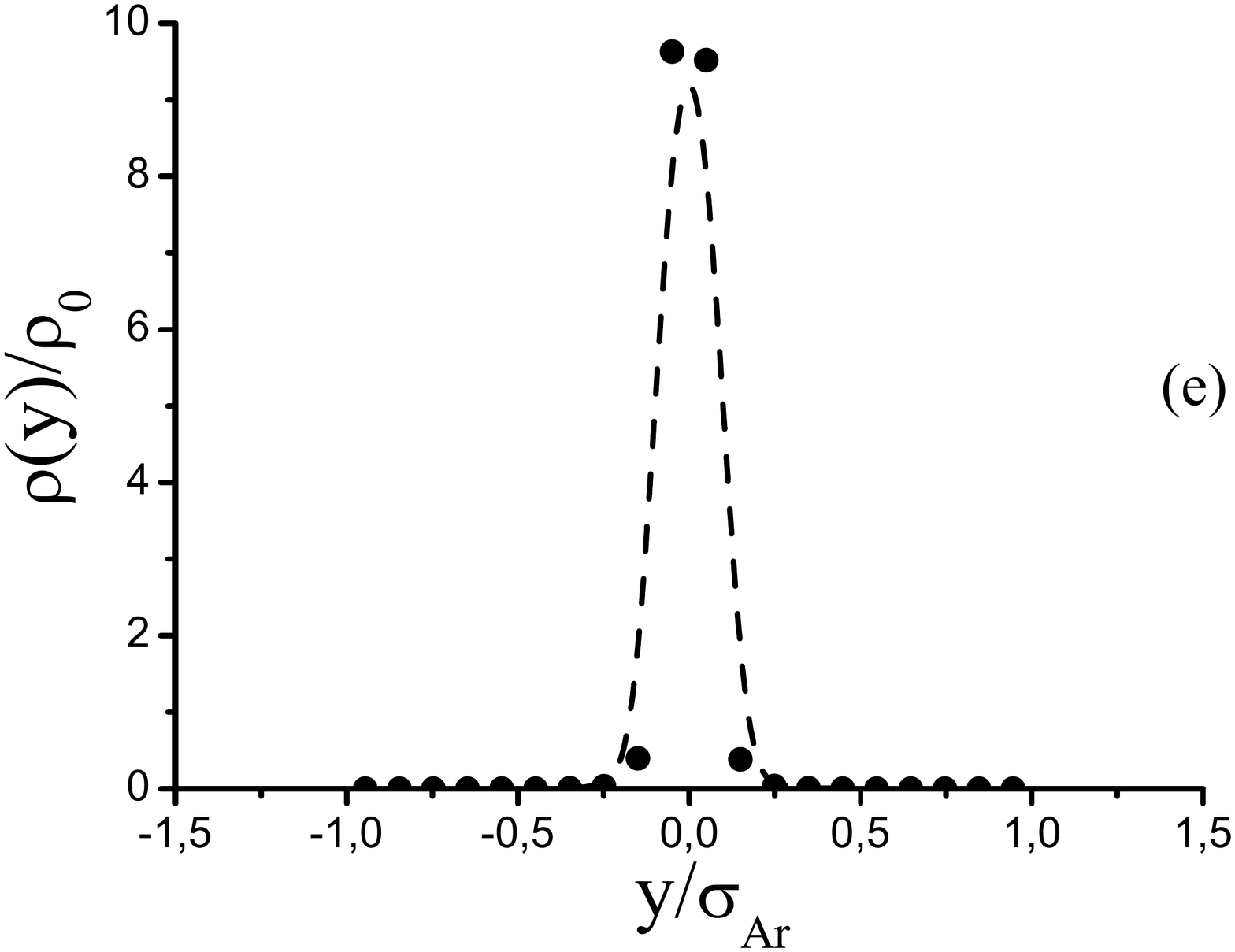}
\caption{\label{f05} Equilibrium density profiles for argon atoms inside SWCNTs with different rectangular cross sections. a - The density profile along 
$z$ axis for SWCNT with square cross section. $\rho_0 = 0.53368$.  b - The analogous profile for SWCNT with rectangular cross section having the ratio 
between its sides 1 : 2. c - The density profile along $y$ axis for the same SWCNT. $\rho_0 = 0.52252$. d - The density profile along $z$ axis for SWCNT 
with rectangular cross section having the ratio between its sides 1 : 4. e - The density profile along $y$ axis for the same SWCNT. $\rho_0 = 0.41445$.}
\end{figure*}

\begin{figure*}[t]\centering
\includegraphics[width = 8cm]{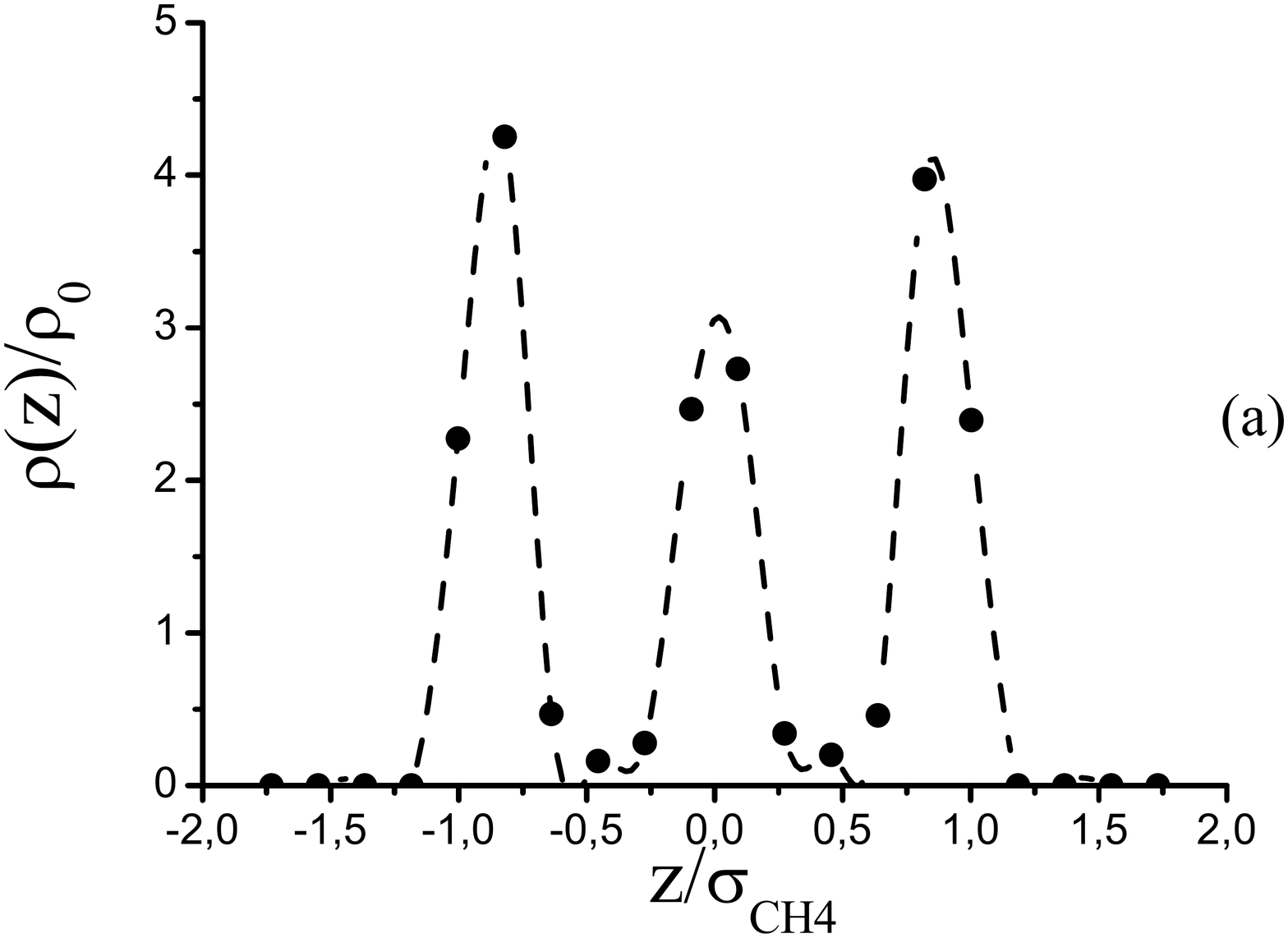}\kern6mm\includegraphics[width = 8cm]{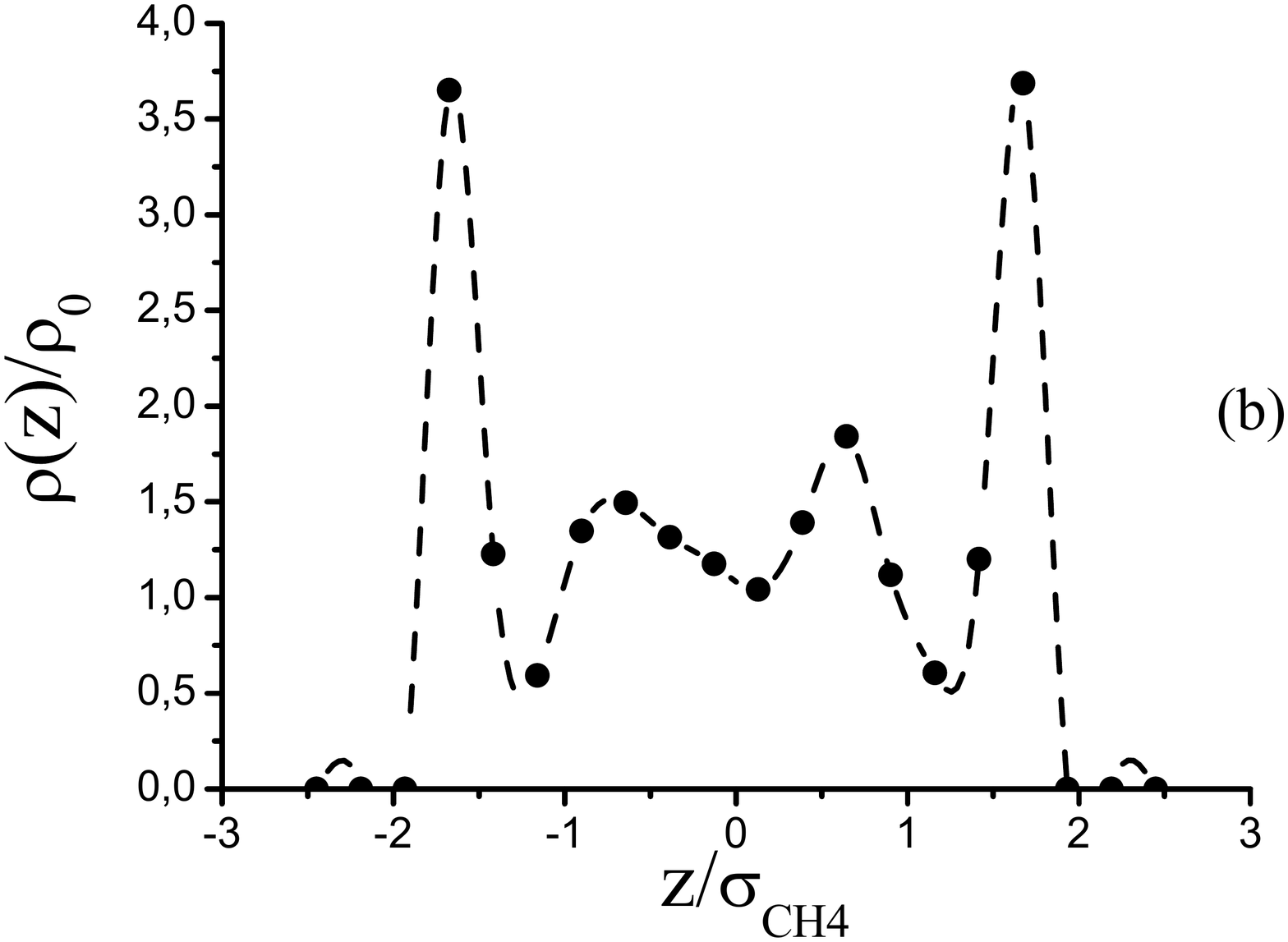}\\ 
\includegraphics[width = 8cm]{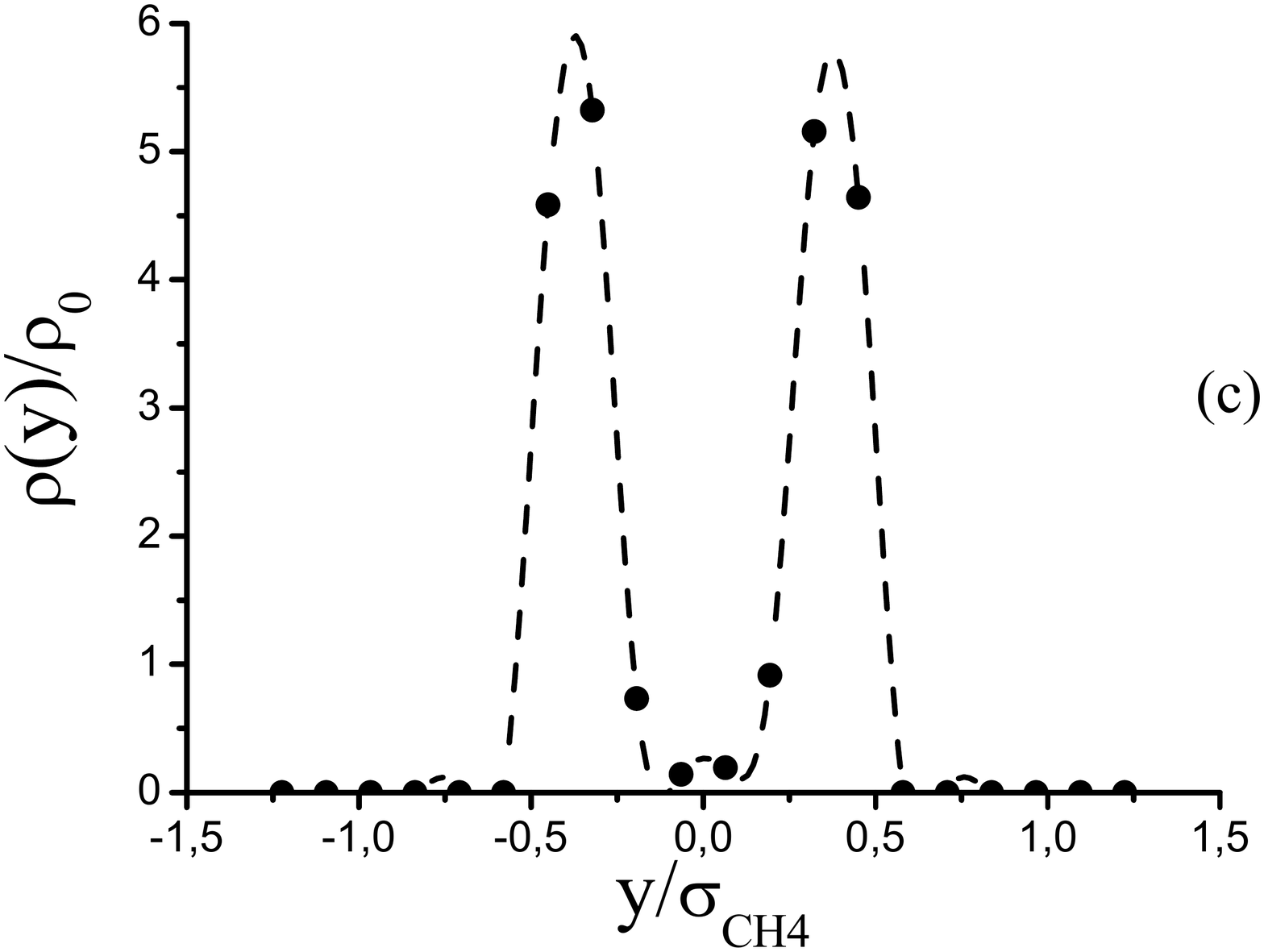}\kern6mm\includegraphics[width  = 8cm]{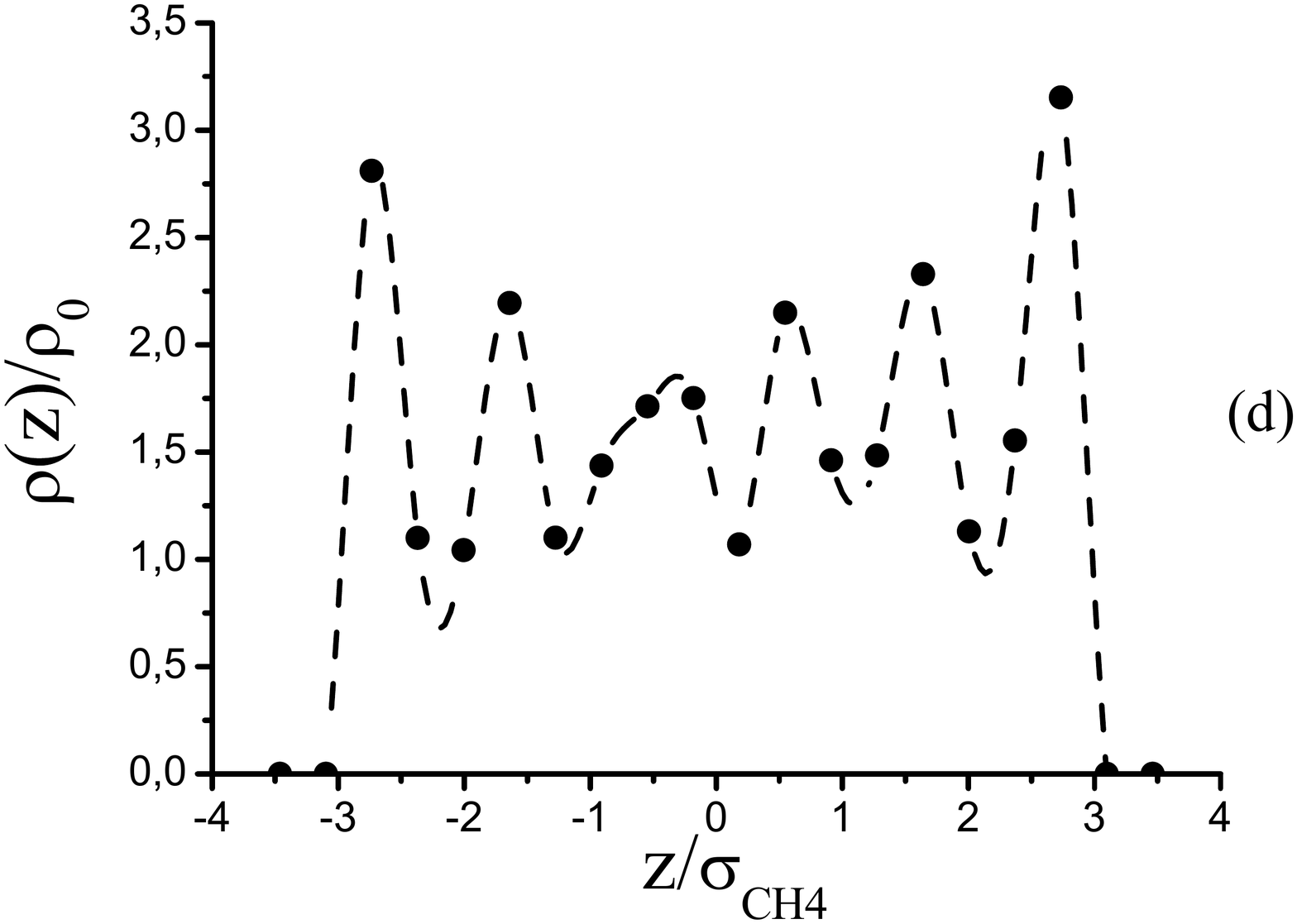}\\
\includegraphics[width = 8cm]{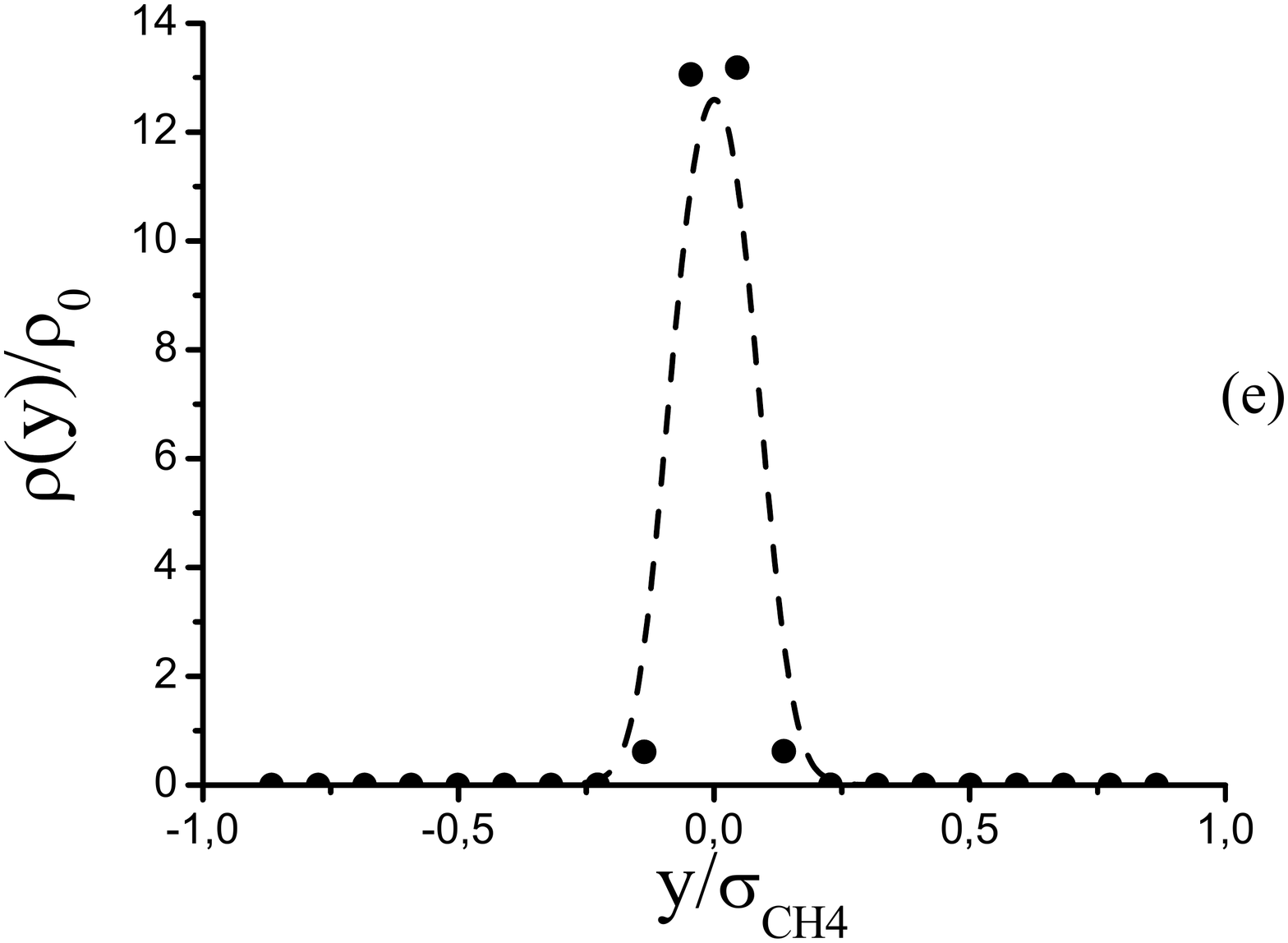}
\caption{\label{f06} Equilibrium density profiles for methane molecules inside SWCNTs with different rectangular cross sections. a - The density profile 
along $z$ axis for SWCNT with square cross section. $\rho_0 = 0.51835$.  b - The analogous profile for SWCNT with rectangular cross section having the ratio
between its sides 1 : 2. c - The density profile along $y$ axis for the same SWCNT. $\rho_0 = 0.38506$. d - The density profile along $z$ axis for SWCNT 
with rectangular cross section having the ratio between its sides 1 : 4. e - The density profile along $y$ axis for the same SWCNT. $\rho_0 = 0.25177$.}
\end{figure*}

\begin{figure*}[t]\centering
\includegraphics[width = 8cm]{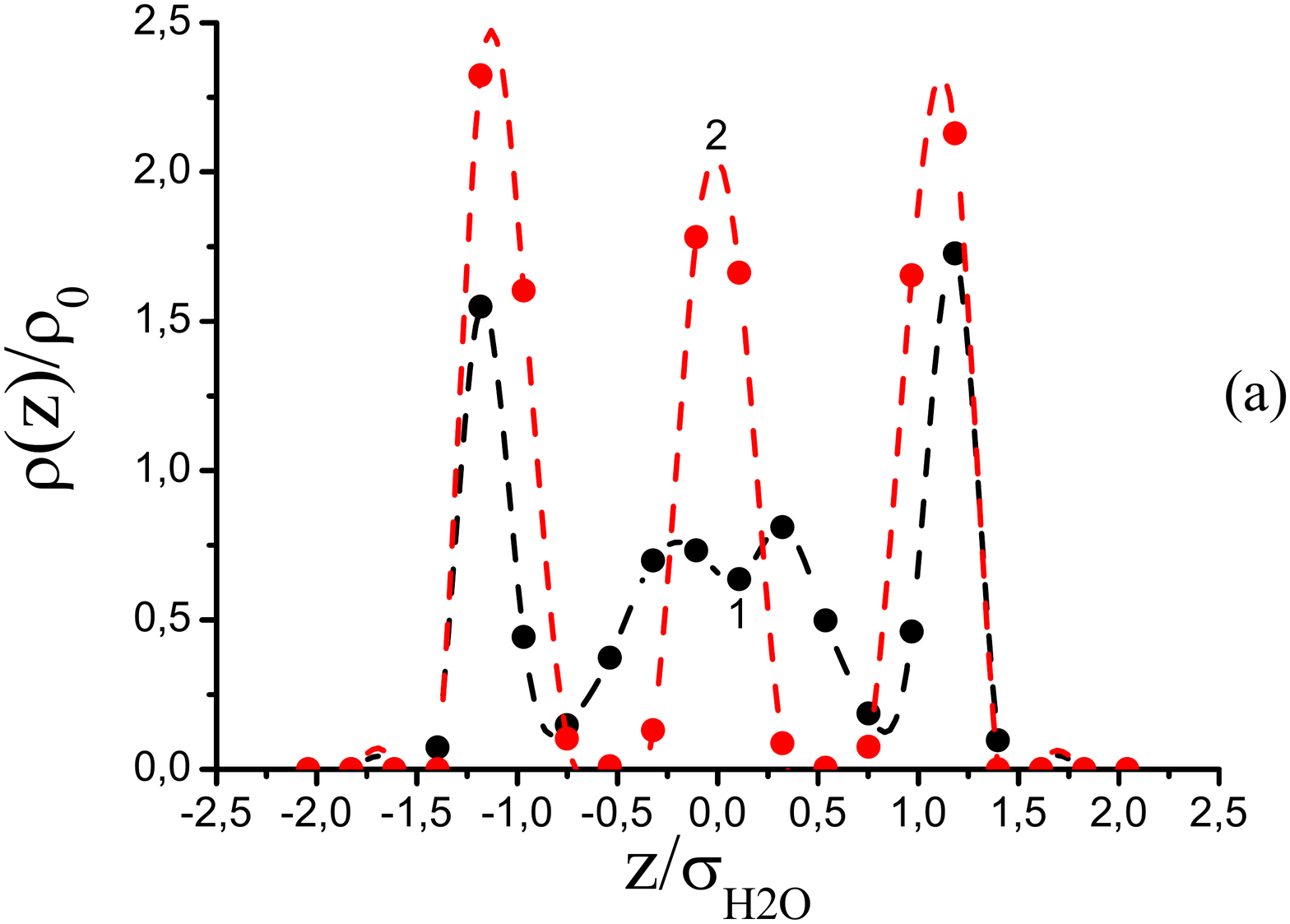}\kern6mm\includegraphics[width = 8cm]{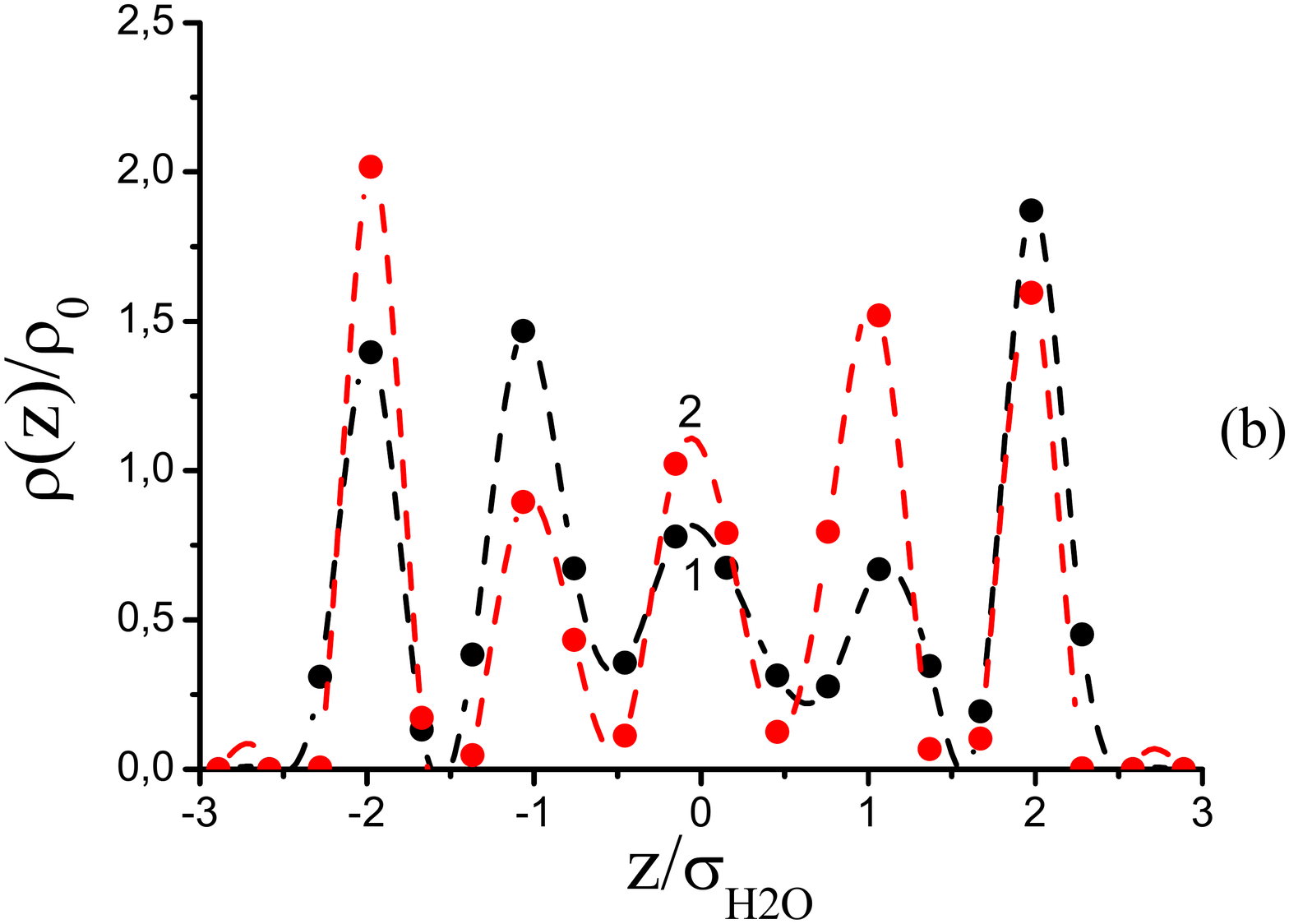}\\ 
\includegraphics[width = 8cm]{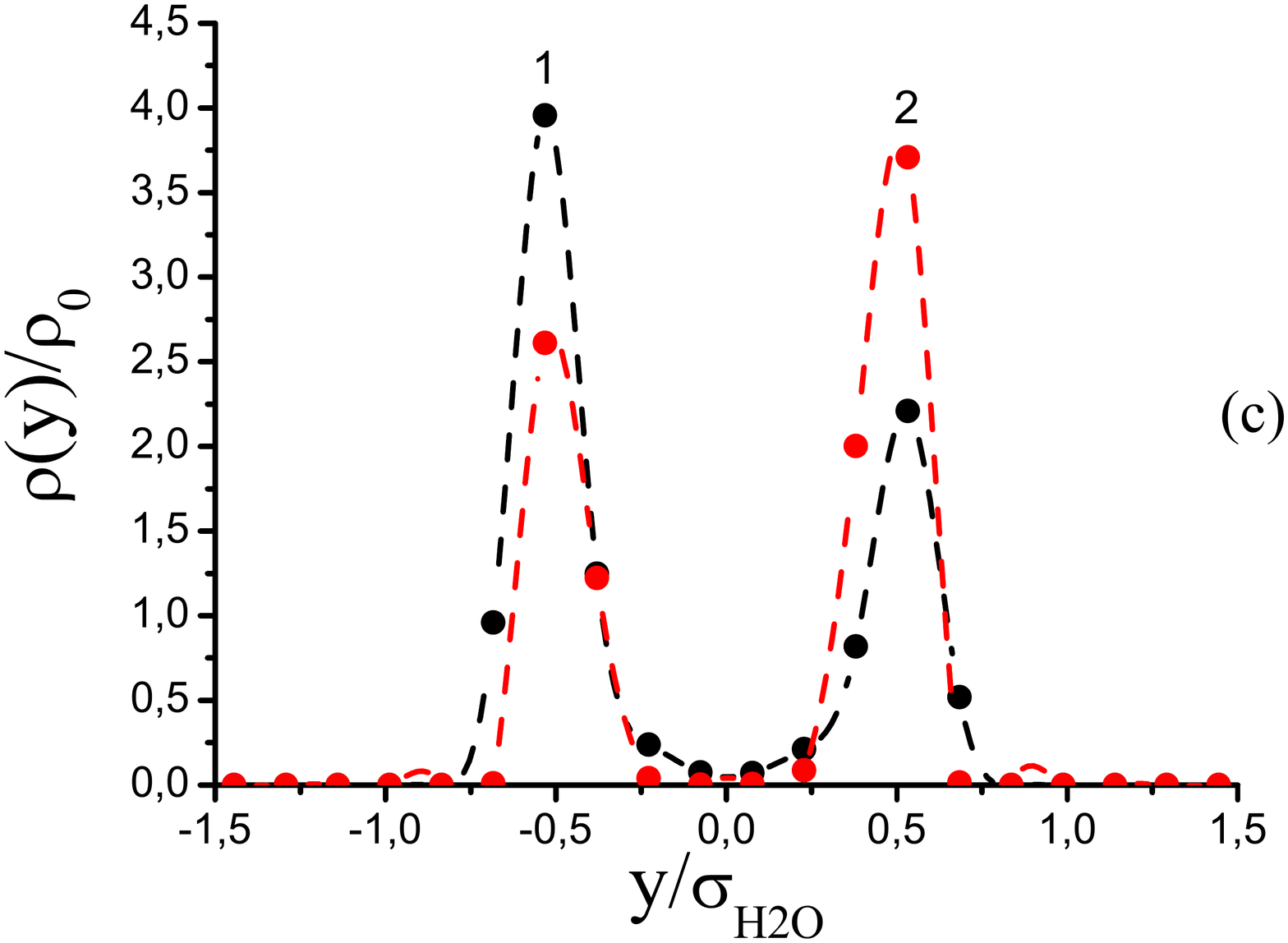}\kern6mm\includegraphics[width = 8cm]{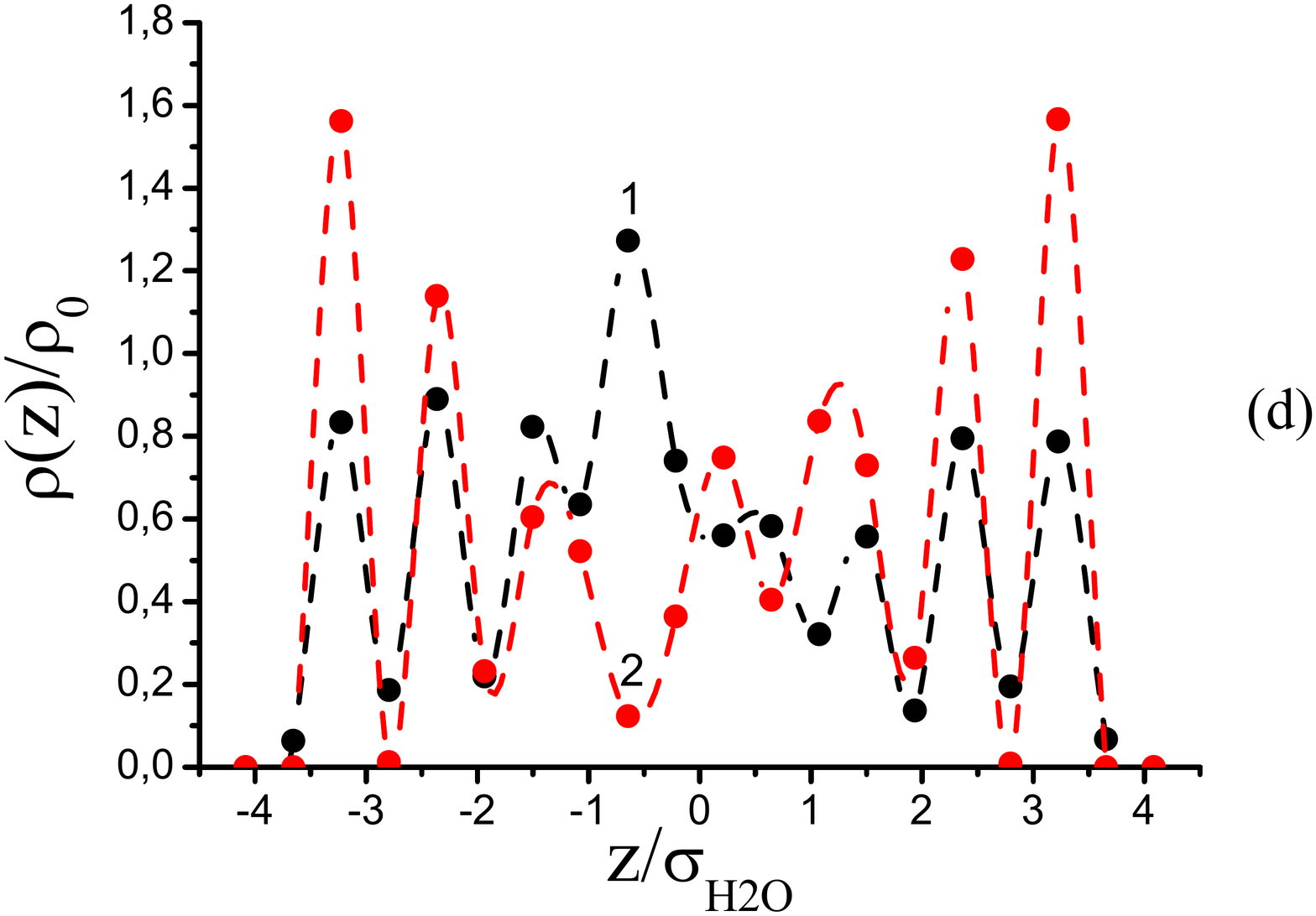}\\
\includegraphics[width = 8cm]{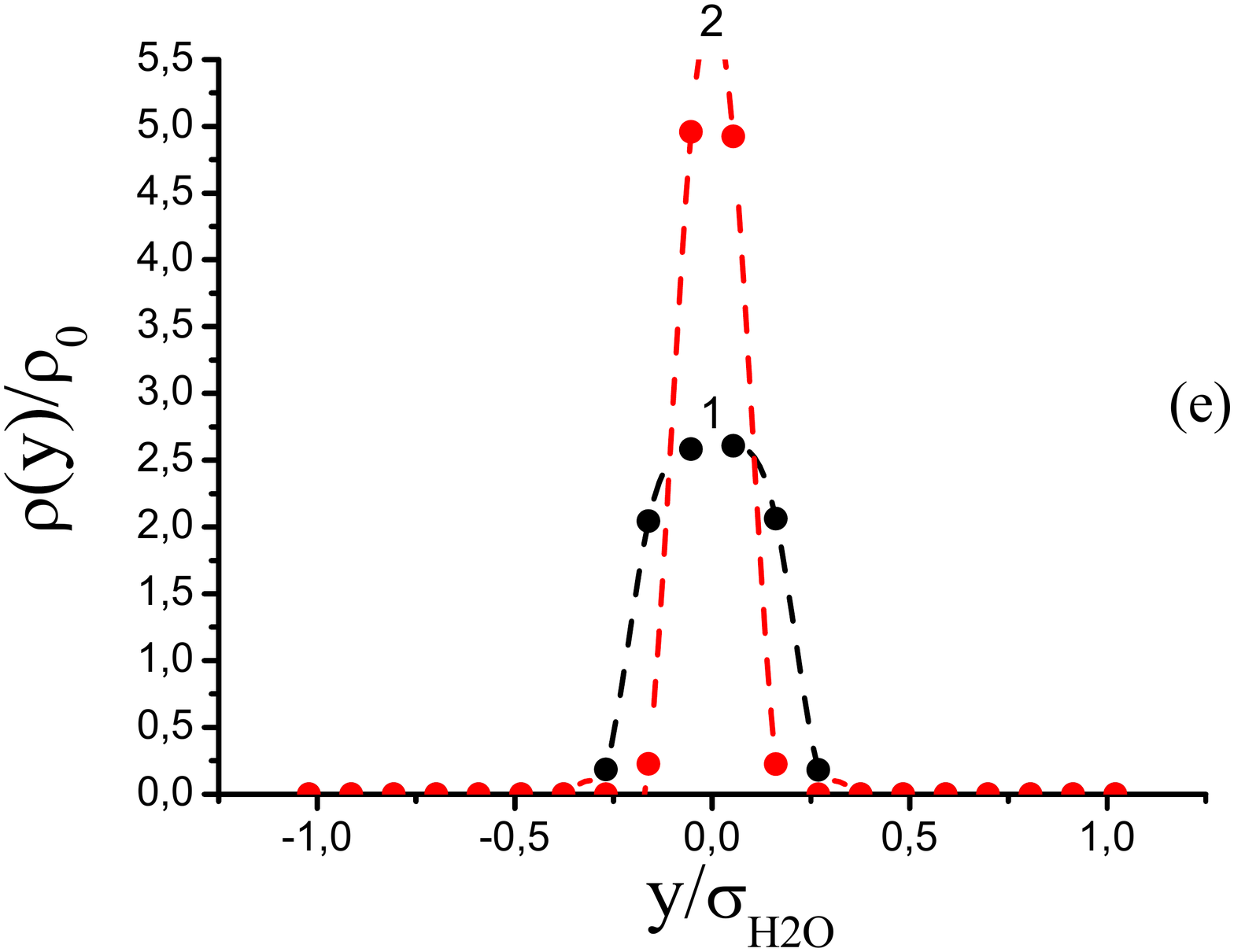}
\caption{\label{f07} Equilibrium density profiles for mixtures of water and methane  molecules inside SWCNTs with different rectangular cross sections 1 - 
the water profiles, 2 - the methane profiles. a - The density profile along $z$ axis for SWCNT with square cross section. $\rho_0 = 0.4597$.  b - The 
analogous profile for SWCNT with rectangular cross section having the ratio between its sides 1 : 2. c - The density profile along $y$ axis for the same 
SWCNT. $\rho_0 = 0.45357$. d - The density profile along $z$ axis for SWCNT with rectangular cross section having the ratio between its sides 1 : 4. 
e - The density profile along $y$ axis for the same SWCNT. $\rho_0 = 0.40112$.}
\end{figure*}

Now, let us turn to results of MD simulations of flows of the polar water, nonpolar methane and argon, and the mixture of the water and methane through 
the above mentioned SWCNTs with rectangular cross sections under action of the external pressure drops across these nanotubes. To simulate these flows, 
in the equilibrium configurations depicted in figure 2(b), we remove the liquid reservoirs and apply periodic boundary conditions to edges of SWCNTs. 
The external pressure drop across natotubes is mimiced by the external force $f_{x0}$ acting on each liquid particle inside SWCNTs. In order to calculate
velocity profiles, we devide the space inside carbon nanotubes into very thin sublayers parallel to the top and bottom bounding walls and calculate 
average molecular velocities inside these sublayers as a function of $z$ coordinates of their centers. 
The fluid flow velocity
profiles obtained from our simulations are shown in figures 8(a) - 8(c) for the polar water, the nonpolar argon, and the mixture of the water and methane, 
respectively. The profiles for the water and argon flows are obtained for the external force equal to $f_{x0} = 0.05$ (in reduced MD units \cite{48}), and 
for the flow of the mixture of the water and methane $f_{x0} = 0.1$ . The nonpolar methane flow
through all rectangular SWCNTs under consideration at these values of $f_{x0}$ is absent. This case will be discussed below. 

One can see from figures 8a - 8c that fluid flows through SWCNTs with rectangular cross sections depend strongly on both the type of the fluid inside 
the tube and  the shape of its cross section. For example, it is easily seen that, for the polar water, the average flow velocity $v_x^{aver}$ should have
a maximum value for SWCNT having rectangular cross section with the ratio between its side 1 : 4 ($v_x^{aver} = 1.15$ in MD units). The intermediate value 
$v_x^{aver} = 0.97$ corresponds to the water flow through SWCNT having rectangular cross section with the ratio between its side 1 : 2, and the minimum
value $v_x^{aver} = 0.28$ exhibits the water flow through SWCNT with square cross section. For nonpolar argon, the average fluid flow velocity has also 
the minimum value $v_x^{aver} = 0.08$ in the case of SWCNT with square cross section but the results for two SWCNTs with other rectangular cross sections
change places: the maximum average flow velocity $v_x^{aver} = 1.2$ corresponds to SWCNT with the rectangular cross section with the ratio between sides
1 : 2, and the fluid flow through other SWCNT with rectangular cross section has the intermediate average fluid flow velocity $v_x^{aver} = 0.7$. Finally,
for the mixture of polar water and nonpolar methane, we have the maximum value $v_x^{aver} = 0.8$ in the case of SWCNT with the square cross section, 
the minimum value $v_x^{aver} = 0.03$ for SWCNT having the rectangular cross section with the ratio between sides 1 : 2, and the intermediate average fluid
flow velocity $v_x^{aver} = 0.4$ corresponds SWCNT with other rectangular cross section. At first glance, it is not easy to give a plausible explanation of
these somewhat intricate results. However, the following qualitative considerations can be made.  

It is clear that the flow of liquid particles through SWCNTs is governed by the external force $f_{x0}$, which is a given constant, and by certain retarding
forces due to the interactions between liquid atoms (molecules) and bounding wall carbon atoms. It is also clear that the stronger these interactions the 
stronger retarding forces, and, hence, the slower the fluid flow. In our simulations, the interactions between liquid atoms (molecules) and the bounding
wall carbon atoms are modelled by means of the LJ pairwise potentials which are characterized by the above mentioned interaction constants 
$\epsilon_{CH2O}$, $\epsilon_{CCH4}$, $\epsilon_{CAr}$, and characteristic lengths $\sigma_{CH2O}$, $\sigma_{CCH4}$, $\sigma_{CAr}$. It is also well known
that these LJ potentials have the minimum disposed at the distance $r^{*}$ to a given carbon atom equal to $r^{*} = 2^{1/6}\sigma$, and, at this minimum,
the force acting on the liquid particle is equal to zero. When the distance between the liquid particle and the carbon atom is less than $r^{*}$ this force is 
repulsive, and, for distances larger than $r^{*}$ it is attractive. When we study the flow of the same liquid particles through different SWCNTs, the interaction 
constant $\epsilon$ for LJ interactions between liquid particles and bounding wall carbon atoms is the same for all SWCNTs, and only distances between wall
atoms and liquid particles define a difference in their flows. Let us consider the water flows through different rectangular SWCNTs. It iseasy to 
calculate average minimum distances $d^{aver}_{min}$ from water molecules to the bounding wall carbon atoms corresponding to equilibrium structures of water
molecules inside these SWCNTs depicted in figures 3a - 3c. These distances are equal to $1.0223\sigma_{H2O}$ for SWCNT with square cross section, 
$1.03\sigma_{H2O}$ for SWCNT with rectangular cross sectionwith the ratio between its sides 1 : 2, and $1.064\sigma_{H2O}$ for SWCNT with other rectangular
cross section with analogous ratio equal to 1 : 4. The distance $r^*$ for LJ interactions between water molecules and bounding wall carbon atoms is equal
to $r^* = 1.131\sigma_{H2O}$. One can see that, for all these SWCNTs, $d^{aver}_{min}$ is smaller than $r^{*}$, and, therefore, inside these nanotubes, the
forces acting on water molecules from bounding wall carbon atoms are repulsive. In addition, the larger difference between $r^*$ and $d^{aver}_{min}$  the
stronger these forces. Hence, these forces are strongest for SWCNT with square cross section, they are weakest for SWCNT with rectangular cross section 
having the ratio between its sides 1 : 4, and, for other SWCNT with rectangular cross section we have an intermediate value for the force between water 
molecules and bounding wall carbon atoms. So, one can conclude that the water flow through SWCNT with rectangular cross section having the ratio between
its sides 1 : 4 should be fastest, for water flow through SWCNT with square cross section should be slowest, and the water flow through SWCNT with other
rectangular cross section should have intermediate average liquid flow velocity. These speculations are in a qualitative agreement with the velocity 
profiles depicted in figure 8a.  One can repeat such qualitative analysis for the flow of argon atoms through the above mentioned SWCNTs with rectangular 
cross sections. For equilibrium structures of argon atoms inside these nanotubes depicted in figures 3d - 3f, we obtain the values of $d^{aver}_{min}$  
equal to $0.96\sigma_{Ar}$,  $0.98\sigma_{Ar}$, and $0.97\sigma_{Ar}$ for SWCNTs having rectangular cross sections with the ratios between their sides 
equal to 1 : 1, 1 : 2, 1 : 4, respectively, and  $r^{*}$ for LJ interactions between argon and carbon atoms equal to $r^{*} = 1.089\sigma_{Ar}$. Therefore,
inside all these SWCNTs, argon atoms are subjected to repulsive forces from the bounding wall carbon atoms, and these forces are strongest for SWCNT with
square cross section, weakest for SWCNT with rectangular cross section with the ratio between its sides 1 : 2, and they have an intermediate value for 
other SWCNT with rectangular cross section. Then, the argon flow should be fastest for SWCNT having the rectangular cross section with the ratio between 
its sides 1 : 2, the average fluid flow velocity should be lowest for SWCNT with square cross section, and the argon flow through SWCNT having rectangular
cross section with the ratio between its sides 1 : 4 should have an intermediate value of the average fluid flow velocity. The results of this analysis
is also in a qualitative agreement with the fluid flow profiles depicted in figure 8b. 

As said above, for the external forces equal to $f_{x0} = 0.05$ and $f_{x0} = 0.1$, the flow of methane molecules through all SWCNTs under consideration
is absent. Therefore, we increased little by little the external force  $f_{x0}$ and found that there are certain threshold or critical values of $f_{x0}$,
$f_{x0}^c$, above which methane molecules can flow through SWCNTs with rectangular cross sections. These critical values, which can be considered as certain
strengths of breakaway, depend strongly on the shape of SWCNT cross sections. We found that, for SWCNT with the square cross section, $f_{x0}^c = 0.275$ (in
reduced MD units), for SWCNT with the rectangular cross section with the ratio between its sides 1 : 2, $f_{x0}^c = 0.15$, and, for SWCNT having the 
rectangular cross section with the ratio between its sides 1 : 4, $f_{x0}^c = 0.8$. The following questions arise: i) Why the liquid methane flows through 
SWCNTs with rectangular cross sections demonstrate an existence of strengths of breakaway that is absolutely not inherent to flows of ordinary liquids?;
ii) Why we do not observe such strengths of breakaway for the water and argon flows through the same SWCNTs?; iii) How can we explain the above mentioned 
dependence of $f_{x0}^c$ on the shape of the SWCNT cross sections?. The answer to the first question seems to be sufficiently obvious. If we look at 
figures 3a - 3i, which exhibit equilibrium structures of argon atoms and water and methane molecules inside SWCNTs under consideration, we can see an 
occurrence of different types of positional order which is not inherent to an ordinary liquid phase. Thus, fluid atoms (molecules) inside our SWCNTs
form solid -like structures, and, as is well known, the strength of breakaway is a typical phenomenon for sliding a solid along a solid surface. 

The answer
to the second question is also simple enough. The interaction constant $\epsilon_{CCH4}$, which defines a strength of interaction between methane molecules
and bounding wall carbon atoms, is considerably larger than analogous constants $\epsilon_{CH2O}$ and $\epsilon_{CAr}$ which define strengths of interactions 
between bounding wall carbon atoms and water molecules and argon atoms, respectively. In addition, the effective size of methane molecules, $\sigma_{CH4}$, 
is larger than effective sizes of argon atoms $\sigma_{Ar}$ and water molecules $\sigma_{H2O}$. Therefore, the interaction between methane molecules and
bounding wall carbon atoms is significantly stronger than analogous interactions of water molecules and argon atoms. Perhaps, their flows through SWCNTs
under consideration could also exhibit certain strengths of breakaway, but these strengths are much lower than the force $f_{x0}$ used to drive argon 
atoms and water molecules.  
                                  
In order to answer to the third question, we should, as we made above, calculate average minimum distances $d^{aver}_{min}$ between methane molecules
and bounding wall carbon atoms for equilibrium structures formed by methane molecules inside SWCNTs under consideration. Our calculations give                                              
$d^{aver}_{min} = 0.94 \sigma_{CH4}$ for SWCNT with square cross section, $d^{aver}_{min} = 0.95 \sigma_{CH4}$ for SWCNT having rectangular cross section
with the ratio between its sides 1 : 2, and $d^{aver}_{min} = 0.89 \sigma_{CH4}$ for SWCNT with other rectangular cross section. We also obtain 
$r^{*} = 1.055\sigma_{CH4}$ for LJ interactions between methane molecules and bounding wall carbon atoms. Repeating the above reasoning about 
relationship between difference $r^{*} - d^{aver}_{min}$ and the strength of interactions between liquid particles and bounding wall carbon atoms, one 
can conclude that such interaction between methane molecules and carbon atoms should be strongest for SWCNT having rectangular cross section with the 
ratio between its sides 1 : 4, weakest for SWCNT with other rectangular cross section, and intermediate for SWCNT with square cross section. Thus, one can
explain why the strength of breakaway should be highest for SWCNT with rectangular cross section with the ratio between its sides 1 : 4, lowest for SWCNT
with other rectangular cross section, and intermediate for SWCNT with square cross section.

The qualitative explanation of velocity profiles for the flows of the mixture H2O + CH4 through SWCNTs with different rectangular cross sections depicted
in figure 8c can be obtained by means of similar analysis. Since in this mixture, methane molecules are characterized by largest constant 
$\epsilon_{CCH4}$ for LJ interactions of these molecules with bounding wall carbon atoms, they are subjected to strongest retarding forces from bounding 
walls. The stronger these forces the slower the fluid of water and methane molecules through SWCNT and vice versa. Therefore, we should calculate 
$d^{aver}_{min}$ for methane molecules in equilibrium structures formed by the mixture H2O + CH4 inside SWCNTs with different rectangular cross sections
(see figures 3j - 3l) and compare these values with $r^{*}$ for LJ interactions between methane molecules and bounding wall carbon atoms. Such calculations
give  $d^{aver}_{min} = 1.097\sigma_{H2O}$ for SWCNT with the square cross section, $d^{aver}_{min} = 1.069\sigma_{H2O}$ for SWCNT with the rectangular
cross section with the ratio between its sides 1 : 2, and $d^{aver}_{min} = 1.075\sigma_{H2O}$ for SWCNT with other rectangular cross sections. $r^{*}$ for
LJ interactions between methane molecules and carbon atoms is equal to  $r^{*} = 1.089\sigma_{H2O}$. Then, one can see that the value $d^{aver}_{min}$ for
SWCNT with the square cross section is closest to $r^{*}$, $d^{aver}_{min}$ for SWCNT with rectangular cross section with the ratio between sides is most 
different from $r^{*}$, and $d^{aver}_{min}$ for SWCNT with other rectangular cross section has an intermediate value between two above mentioned ones. Then,
one can conclude that the flow of the mixture H2O + CH4 through SWCNT with the square cross section should be fastest, the flow of this mixture through
SWCNT with the rectangular cross section with the ratio between its sides 1 : 2 should be slowest, and the flow of such mixture through SWCNT with other 
rectangular cross section should have an intermediate average flow velocity. It is easily seen than these conclusions are in a qualitative agreement with 
the velocity profiles depicted in figure 8c.

As said above, for external driving force $f_{x0} = 0.05$, argon atoms and water molecules flow through SWCNTs with different rectangular cross sections 
with steady and finite average flow velocities $v_x^{aver}$. It means that, since the external force $f_{x0}$ is switched on, liquid particles begin to move 
along the tube axis with a certain accelerations untill the average fluid flow velocity achieves the steady value $v_x^{aver}$. This is a quite expected  
behavior of fluid flows through SWCNTs. However, when the external driving force is two times larger, $f_{x0} = 0.1$, the situation changes radically, and
one can observe two drastically different behaviors that depend on types of fluid particles and the shapes of rectangular sections of SWCNTs. In the case 
of the water flows through SWCNTs with different rectangular cross sections, one can observe again the flows with steady average flow velocities $v_x^{aver}$,
which are higher than those for $f_{x0} = 0.05$, but remain finite. For argon atom flows through SWCNTs with square cross section and rectangular cross
section with the ratio between its sides 1 : 2, one can observe the similar fluid flows with steady and finite flow velocities (see curves 1 and 2 in figure
9a). This figure exhibits time dependences of $v_x^{aver}$ averaged over subsequent time intervals with a duration equal to 100 MD time units (symbols on 
these curves correspond to central points of such time intervals). One can see that, for argon flows through SWCNTs with such rectangular cross sections,
the fluid flow velocities averaged over subsequent time intervals first grow with time, and then reach saturation at certain steady and finite values.
However, for argon flow through SWCNT with the rectangular cross section with the ratio between its sides 1 : 4, the fluid flow velocity averaged over 
above mentioned subsequent time intervals exhibit an unlimited growth with no signs of saturation. Moreover, if we then switch off the external force 
($f_{x0} = 0$, the average flow velocity remains constant with no signs of decay (curve 4 in figure 9a). 
In order to understand such extraordinary behavior 
of argon flows through SWCNTs with different rectangular cross sections, we must analyze a time dependence of all forces acting on argon atoms during
their flows through SWCNTs. Each atom (molecule) inside SWCNT is subjected to two forces directed along the tube axis, namely, the external driving force
$f_{x0}$ and retarding force $f_{rx}$ due to interactions between a given atom (molecule) and bounding wall carbon atoms. The external driving force 
$f_{x0}$ is constant, and typical time dependence of instant value of $f_{rx}$ is shown in figure 9b. It is easily seen that this time dependence has a 
stochastic - like character, and we must perform time averaging of this force over the above mentioned subsequent time intervals. The results of such time 
averaging for argon flows through SWCNTs with different rectangular cross sections are shown in figure 9c, which exhibits time dependences of the ratio
$|f_{rx}|/f_{x0}$, where $|f_{rx}|$ is the absolute value of the time averaged retarding force $f_{rx}$ (if $f_{x0}$  is positive $f_{rx}$  after time
averaging is always negative) for argon flows through SWCNTs with different rectangular cross sections. One can see from this figure that for argon flows
through SWCNTs with square cross section and rectangular cross section with the ratio between its sides 1 : 2, the ratios $|f_{rx}|/f_{x0}$ first grow with
time and then reach saturation at the steady value equal to nealy 1 (curves 1 and 2). It means that the absolute value of the time averaged retarding force $f_{rx}$ 
becomes equal to the external driving force $f_{x0}$, but it has an opposite sign. As a result, the total force acting on carbon atoms during their flows
through SWCNTs vanishes, and they move with certain constant time averaged velocities. One can also see from this figure (curve 3) that, for argon flow 
through SWCNT with rectangular cross section having the ratio between its sides 1 : 4, the ratio $|f_{rx}|/f_{x0}$ decays with time to nearly zero, and 
argon atoms begin to move along tube axis in a ballistic regime. This fact can explain the unlimited growth of the average fluid velocity during argon flow 
through SWCNT with such rectangular cross section under action of the external driving force $f_{x0} = 0.1$. It should be noted that qualitatively  similar
phenomenon, namely, ballistic frictionless gas flow through two - dimensional channels made from graphene or boron nitride has been experimentally observed 
\cite{53}. 

\begin{figure*}[t]\centering
	\includegraphics[width = 8cm]{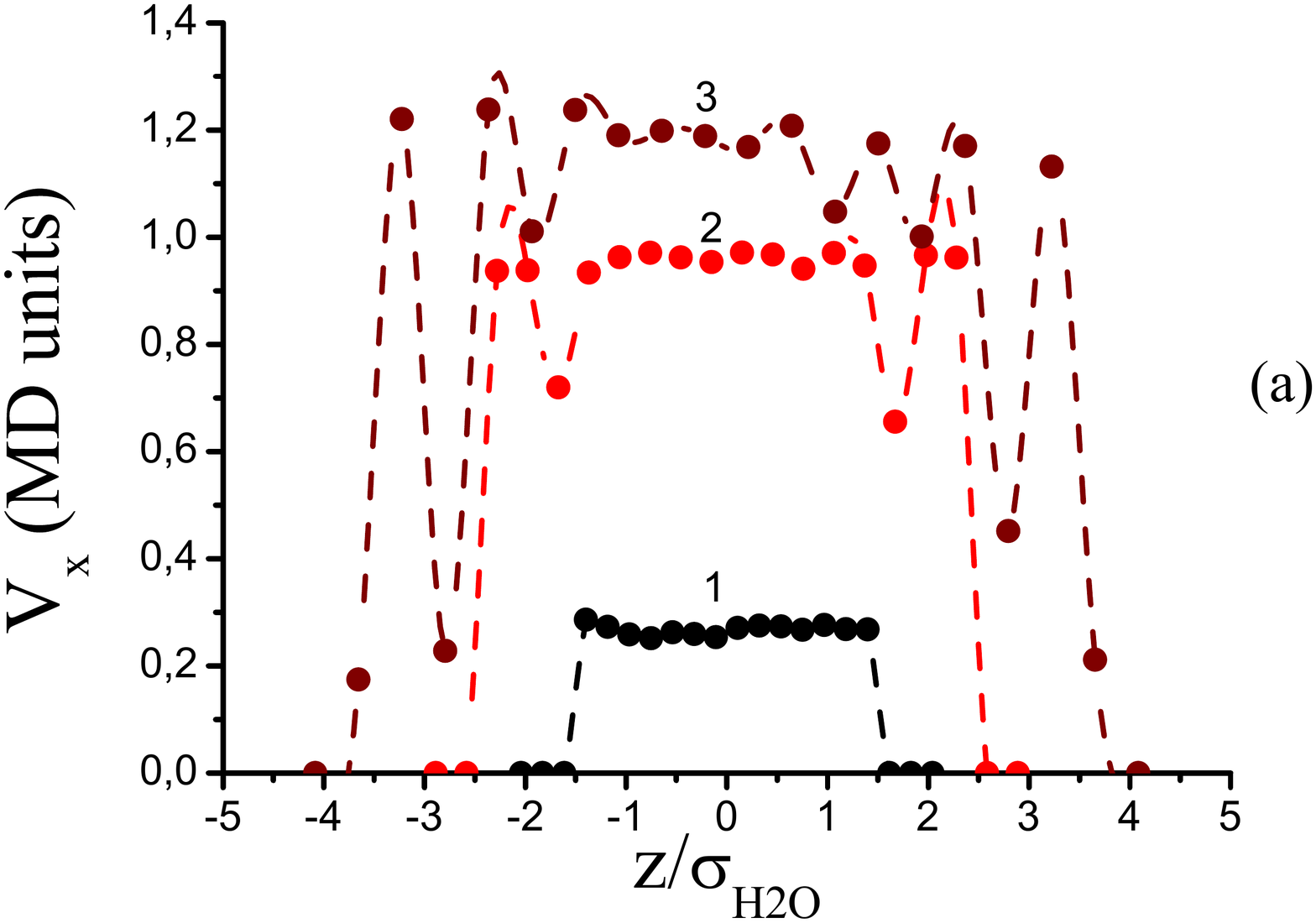}\kern6mm\includegraphics[width = 8cm]{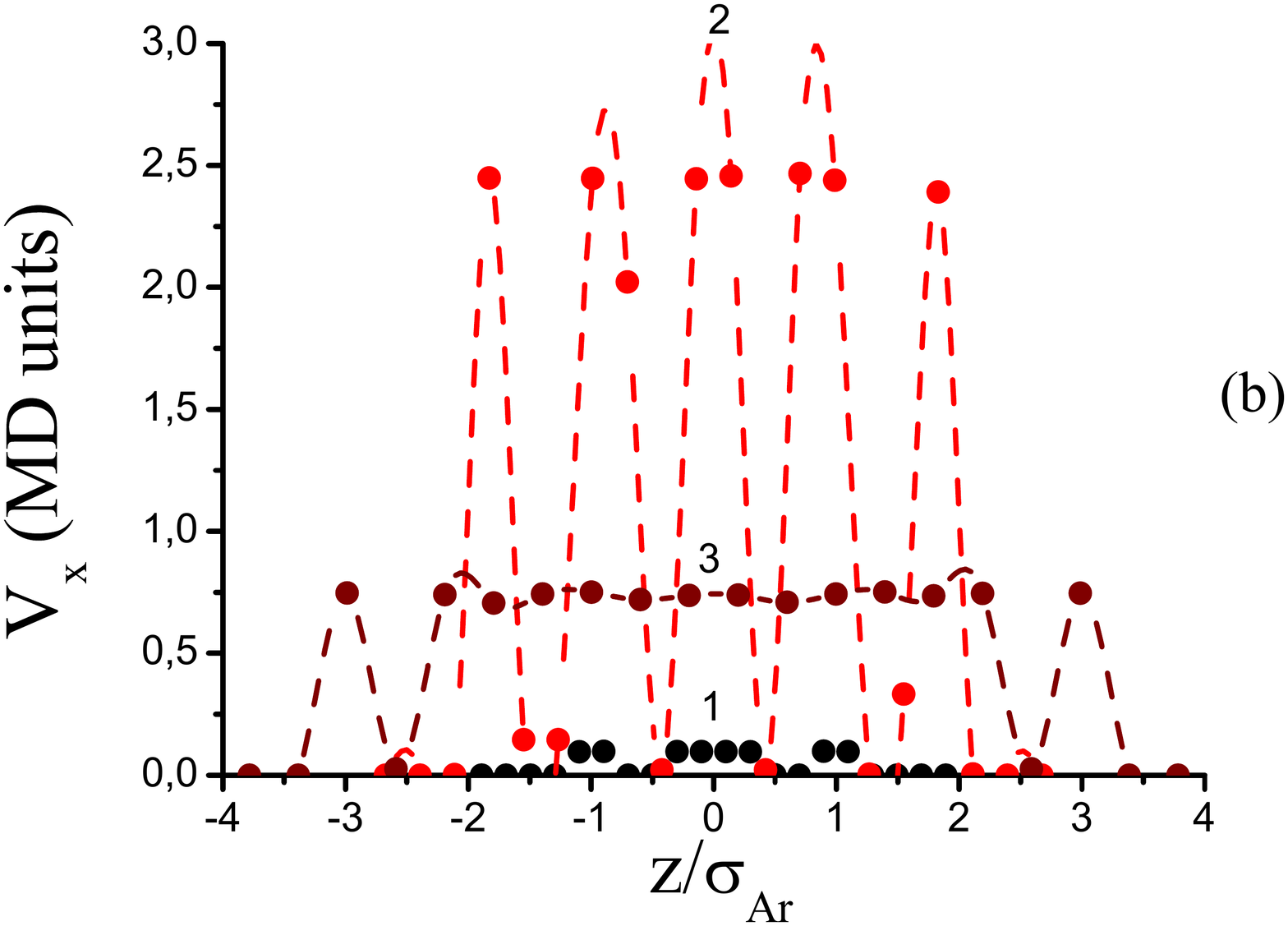}\\
	\includegraphics[width = 8cm]{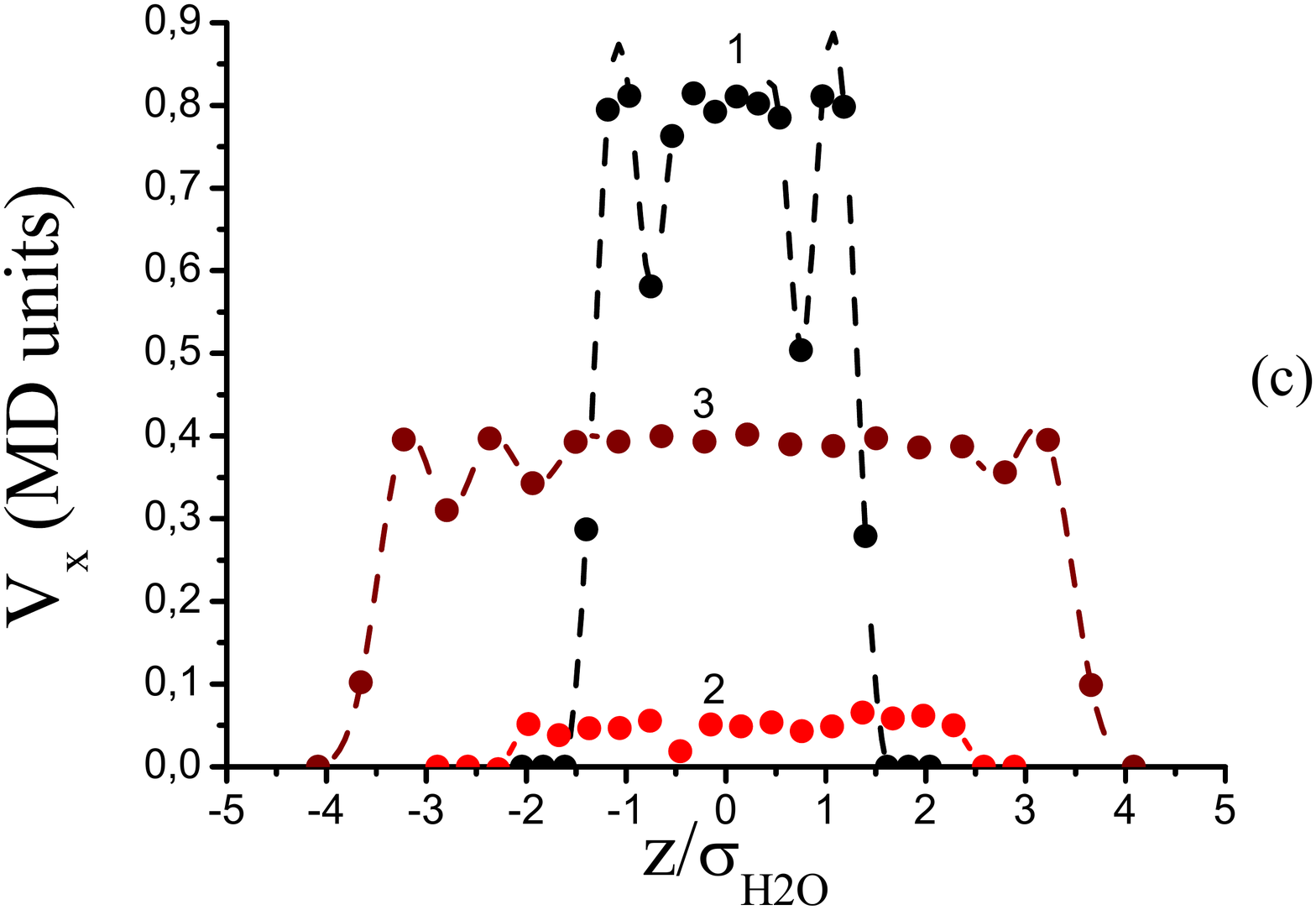}
	\caption{\label{f08} The fluid flow velocity profiles for flows of polar water molecules, nonpolar argon atoms and mixtures of water and methane molecules
		through SWCNTs with different rectangular cross sections. 8a - fluid flow velocity profiles for water molecules, 8b and 8c - analogous profiles for argon atoms 
		and mixtures of water and methane molecules, respectively. $ f_{x0} = 0.05$ for 8a and 8b, and $ f_{x0} = 0.1$ for 8c. Curves 1 in all figures correspond to 
		SWCNT with square cross section; curves 2 and 3 correspond to SWCNTs with rectangular cross sections having the ratio between their sides 1 : 2 and
		1 : 4, respectively.}
\end{figure*}

\begin{figure*}[t]\centering
	\includegraphics[width = 8cm]{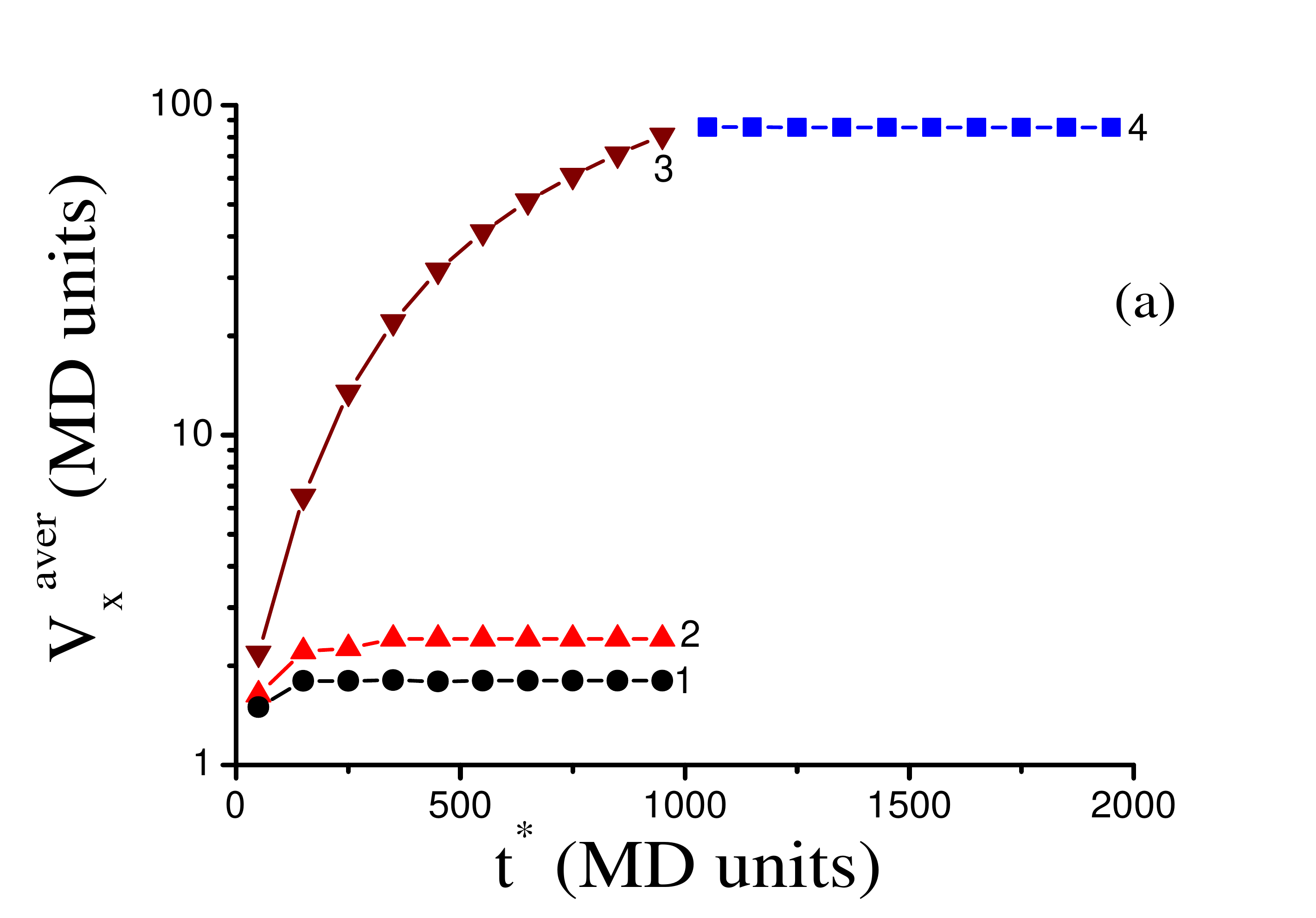}\kern6mm\includegraphics[width = 8cm]{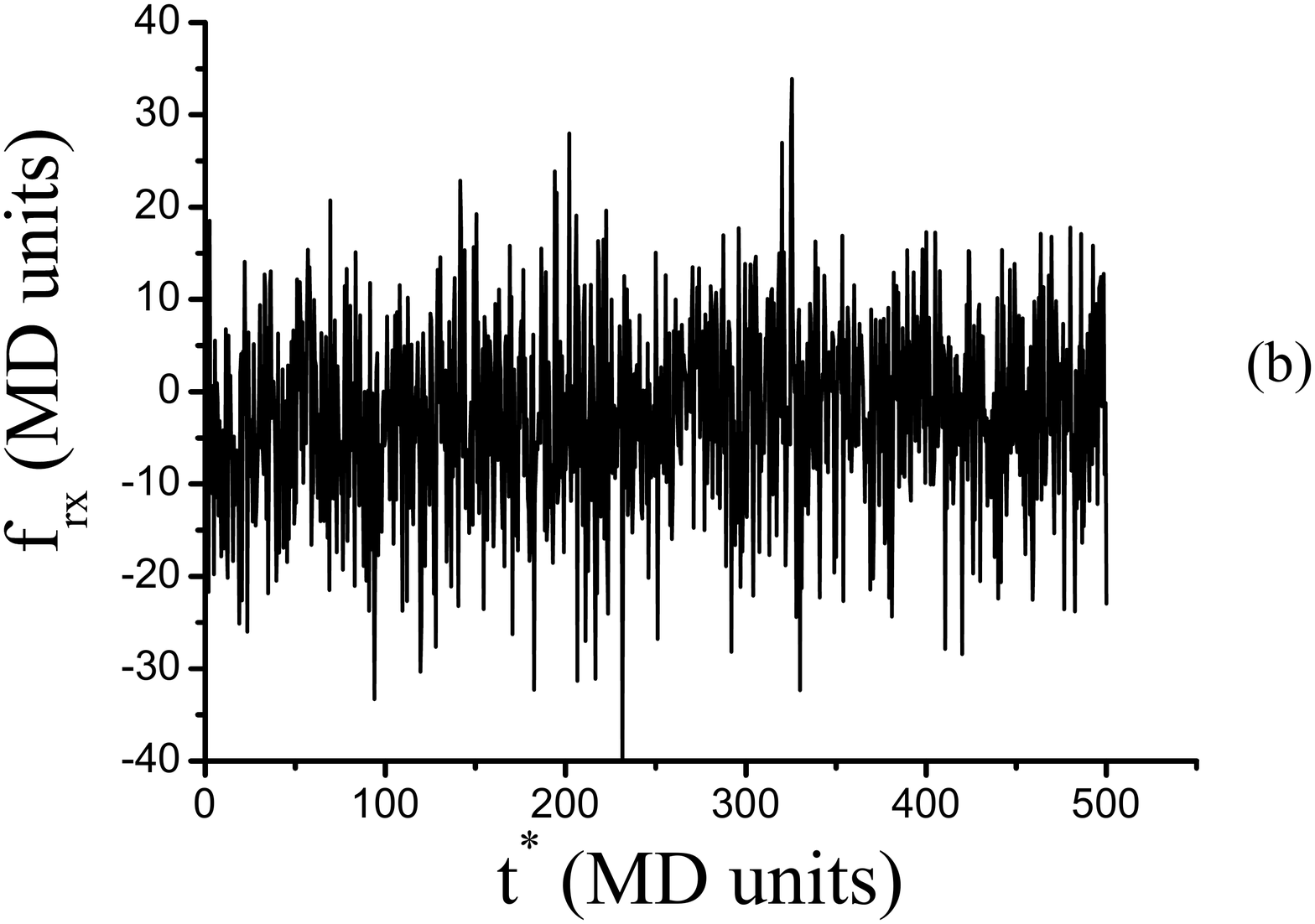}\\ 
	\includegraphics[width = 8cm]{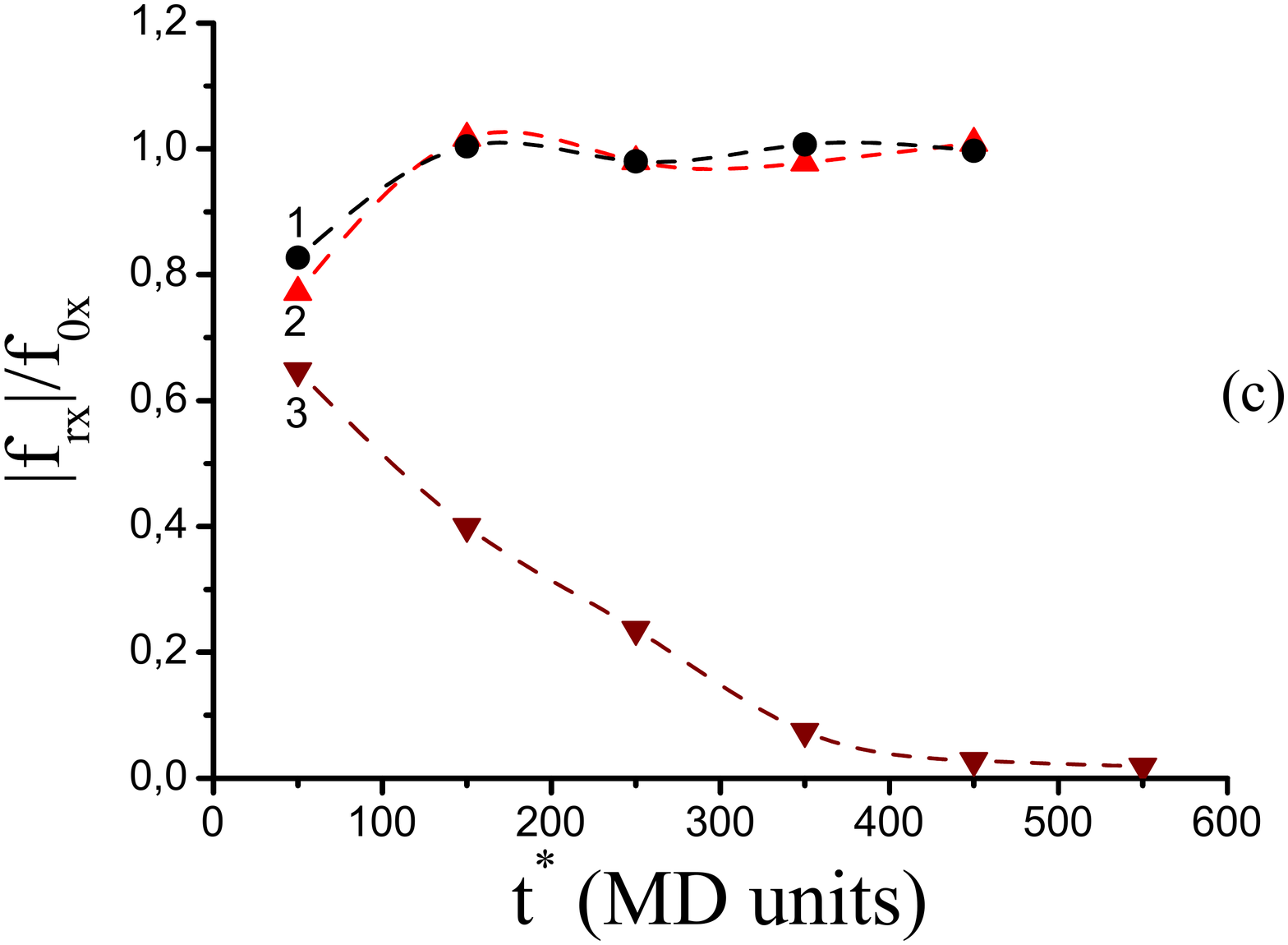}
	\caption{\label{f09} Time dependences of average argon flow velocities $v_x^{aver}$  through SWCNTs with different rectangular cross sections and
		analogous dependeces for retarding forces acting on argon atoms from bounding wall carbon ones. 9a - time dependence of $v_x^{aver}$  averaged over 
		subsequent time intervals with duration equal to 100 reduced MD units. 9b - typical time dependence of instant retarding force $f_{rx}$ acting on
		argon atoms from bounding wall carbon atoms during argon flow through SWCNT with rectangular cross section with the ratio between its sides 
		1 : 4. 9c - time dependences of the ratios $|f_{rx}|/f_{x0}$ averaged over above subsequent time intervals during the argon atom flows through SWCNTs
		with different rectangular cross sections. In 8a and 8c curves 1 correspond to SWCNT with square cross section; curves 2 and 3 correspond to SWCNTs 
		with rectangular cross sections having the ratios between their sides 1 : 2 and 1 : 4, respectively. For all figures $f_{x0} = 0.1$.}
\end{figure*}

At first glance, this phenomenon seems to be somewhat similar to the superfluidity that occurs, for example, in helium - 4 near the absolute zero 
\cite{54}. However, there are several principal differences between well known regular classic superfluidity and our results on argon flow through SWCNT 
with one of above mentioned rectangular cross sections. First of all, the regular classic superfluidity is the macroscopic quantum phenomenon whereas 
our MD simulations are based on the usual classic mechanics. Secondly, our "pseudo superfluidity" depends on the shape of the cross section of SWCNT, 
whereas the "true" superfluidity is independent of shapes of channels. Thirdly, when the external driving force is equal to $f_{x0} = 0.05$, the average
fluid argon flow velocity through our SWCNT is finite, whereas disappearance of viscosity of helium - 4 depends only on its temperature and does not depend 
on the external driving forces. Thus, the results of our simulations on the liquid argon flows through SWCNTs with rectangular cross sections have nothing
to do with the classic superfluidity. Perhaps, these results are due a combination of several factors, namely, the equilibrium structure formed by argon 
atoms inside SWCNT with the rectangular cross section with the ratio between its sides 1 : 4, and the time averaging of retarding forces acting on argon 
atoms from bounding wall carbon atoms. May be, the analysis of the time dependence of the instant retarding force $f_{rx}$ depicted in figure 9b will allow
us to elucidate this challenge.

\section{Conclusion}\label{sec:conclusion}

We performed MD simulations of  equilibrium structures and flows of polar water, nonpolar argon and methane, and mixtures of water and methane confined 
by SWCNTs with square cross section and rectangular cross sections having the same area and the ratios between their sides 1 : 2 and 1 : 4. The results 
of our simulations show that equilibrium structures of all confined liquids depend mainly on the shape of the SWCNT's rectangular cross sections, namely, 
the cross sections of these structures resemble replicas of those of SWCNTs. Nevertheless, the types of liquids confined by above mentioned SWCNTs also
have some influence on their equilibrium structures. For example, the results of performed MD simulations revealed that nonpolar argon atoms form inside 
SWCNTs with rectangular cross sections the most spatially ordered equilibrium structures, whereas, polar water molecules form the least spatially ordered
ones. The corresponding decrease in the spatial order is due to the Coulomb - like dipole - dipole interactions between polar water molecules. As for 
the external pressure driven flows of all above mentioned liquids through SWCNTs with different rectangular cross sections, these flows depend strongly on 
both the shapes of the rectangular cross sections and the type of the confined liquids. For example, our MD simulations revealed that, for nonpolar methane
inside above SWCNTs  with different rectangular cross sections, there are critical (threshold) values of the external driving force $f_{x0}$, that mimics
the external pressure drop through SWCNTs, below which the average flow velocity is nearly zero, and above which the liquid methane flow occurs. Our 
simulations revealed also that these critical values, which can be considered as certain strengths of breakaway, depend strongly on the shape of rectangular 
cross sections of our SWCNTs. Perhaps, this phenomenon, which is absolutely not inherent to flows of ordinary liquids, is due to a certain spatial order 
formed by argon and methane atoms inside SWCNTs with rectangular cross sections. We show that observed dependence of the strengths of breakaway obtained 
from our MD simulations on the shapes of rectangular cross sections of SWCNTs can be qualitatively explained in terms of interactions between liquid 
particles and bounding wall carbon atoms. The stronger these interactions the stronger the strength of breakaway for SWCNTs with rectangular cross sections.

Another interesting phenomenon was revealed from our MD simulations of the liquid argon flows through above mentioned SWCNTs with rectangular cross
sections. For the external driving force $f_{x0} = 0.05$, these flows are characterized by the average flow velocities $v_x^{aver}$ that are different for
different shapes of SWCNT's rectangular cross sections but still remain finite. However, for $f_{x0} = 0.1$, the liquid argon flows through SWCNTs with 
square cross section and rectangular cross section with the ratio between its sides 1 : 2 also exhibit different finite values of $v_x^{aver}$, whereas
the liquid argon flow through SWCNT with rectangular cross section with the ratio between its sides 1 : 4 occurs in the ballistic regime, i. e., the 
average flow velocity $v_x^{aver}$ exhibits an unlimited growth with time. It was revealed that, for the liquid argon flows through two former SWCNTs 
with rectangular cross sections, the retarding force $f_{rx}$ acting on liquid particles from the bounding wall carbon atoms averaged over a certain 
subsequent time intervals first grows with time until it reaches a saturation at the value equal to the external driving force $f_{x0} = 0.1$ but with
an opposite sign. For the argon flow through SWCNT with rectangular cross section with the ratio between its sides 1 : 4, the same time averaged retarding 
force $f_{rx}$ decreases with time to nearly zero. Therefore, for the liquid argon flows through two former SWCNTs with rectangular cross sections, the
total steady force acting on argon atoms along the tube axis is equal to zero, whereas, for liquid argon flow through SWCNT with rectangular cross section
with the ratio between its sides 1 : 4, this force is almost equal to the external driving force $f_{x0} = 0.1$. This fact can explain a difference in 
behaviors of the liquid argon flows through above mentioned SWCNTs with different rectangular cross sections.


\begin{thebibliography}{54}

\bibitem{01} S. Iijima. Nature. {\bf 354}, (1991), 56. 
\bi{02}  B. I. Yakobson and E. S. Richard.  American Scientist.{\bf 85}, (1997), 324. 
\bi{03} E. T. Thostenson,  Z. F. Ren, and T .W. Chou. Compos. Sci. Technol. {\bf 61}, (2001), 1899. 
\bi{04} D. Qian, G. F. Wagner, W. K. Liu, M. F. Yu, and R. S. Ruoff. Appl. Mech. Rev. {\bf 55}, (2002), 495. 
\bi{05} C. Y. Wang, Y. Y. Zhang, C. M. Wang, and V. B. C. Tan. J. Nanosci. Nanotech. {\bf 7}, (2007), 4221.
\bi{06} D. S. Bethune and C. H. Kiang. Nature. {\bf 363}, (1993), 605. 
\bi{07} M. S. Dresselhaus, G. Dresselhaus, and R. Saito. Carbon {\bf 33},(1995), 883. 
\bi{08} W. E. Thomas. Phys. Today. {\bf 49}, (1996), 26. 
\bi{09} M. S. Dresselhaus, G. Dresselhaus, and P. Avouris. \emph {Carbon nanotubes}.Clarendon Press, Oxford, 1989.
\bi{10} P. G. Collins and P. Avouris. Scientific. American. {\bf 283},(2000), 62. 
\bi{11} J. K. Holt, H. P. Park, Y. Waang, M. Staderman, A. B. Artyukhin, C. P. Grigopopulos,
A. Noy, and O. Bakajin, Science {\bf 312}, (2006), 1034. 
\bi{12} M. Majumder, N. Chopra, R. Andrews, and B. J. Hinds. Nature {\bf 438}, (2005), 44. 
\bi{13} M. Majumder, N. Choudhury, and S. K. Ghosh. J. Chem. Phys. {\bf 127}, (2007), 054706. 
\bi{14} N. Choudhury and B. M. Pettitt. J. Phys. Chem. {\bf 109}, (2005), 6422. 
\bi{15} B. Mukherjee, P. K. Maiti, C. Dasgupta, and A. K. Sood. J. Chem. Phys. {\bf 126}, (2007), 124704. 
\bi{16} M. Melillo, F. Zhu, M. A. Snyder, and J. Mittal. J. Phys. Chem. Lett. {\bf 2}, (2011), 2978. 
\bi{17} N. Chopra and N. Choudhury. J. Phys. Chem. C.{\bf 117}, (2013), 18398.
\bi{18} S. K. Kannam, B. D. Todd, J. S. Hansen, P. Davis. J. Chem. Phys. {\bf 138}, (2013), 094701.
\bi{19} J. Su, K. Yang. Chemphyschem. {\bf 16}, (2015), 3488.
\bi{20} X. Meng and J. Huang. Molecular Simulation. {\bf 42}, (2015), 215.
\bi{21} A. Sam, S. K. Kannam, R. Hartkamp, S. P. Sathian. J. Chem. Phys. {\bf 146}, (2017), 234701.
\bi{22} M. E. Suk and N. R. Aluru. Nano. Micro. Thermophys. Eng. {\bf 21}, (2017), 247. 
\bi{23} S. J. Klaine, P. J. J. Alvarez, G.E. Batley, T. F. Fernandes, R. D. Handy, D. Y. Lyon, S. Mahendra, M. J. McLaughlin, and J. R. Lead. 
Environ. Phys. Toxicol. Chem.  {\bf 29}, (2008), 1825.
\bi{24} M. S. Mauter and M. Elimelech. Sci. Technol. {\bf 42}, (2008), 5843. 
\bi{25} A. Keller, S. McFerran, A. Lazareva, S. Suh. J. Nanoparticles. Res. {\bf 15}, (2013), 1.
\bi{26} V. K. K. Upadhyayula, S. Deng, M. C. Mitchell, and G. B. Smith. Sci. Notal. Environ. {\bf 408}, (2009), 1. 
\bi{27} S. Zhang, T. Shao, and S. S. K. Bekaroblu, T. Karanfil. Water. Res. {\bf 45}, (2011), 1378. 
\bi{28} O. G. Opul and T. Karanfil. Water. Res. {\bf 68}, (2015), 34. 
\bi{29} D. B. Geohegan, A. A. Puretzky, I. N. Ivanov, S. Jesse, G. Eres,  and J. Y. Hove. Appl. Phys. Lett. {\bf 83}, (2003), 1851. 
\bi{30} L. C. Venema, J. W. G. Wildoer, H. L. J. T. Tuinstra, C. C. Dekker, A. G. Rinzler, and R. E. Smalley. Appl. Phys. Lett. {\bf 71}, (1997), 2629. 
\bi{31} S. Iijima and T. Ichihashi. Nature. {\bf 363}, (1993), 603. 
\bi{32} D. S. Bethune, C. H. Kiang, M. S. D. Vries, G. Gorman, R. Savoy, J. Vazquez, and R. Beyers. Nature. {\bf 363}, (1993), 605. 
\bi{33} T. Hiraoka, S. Bundow, H. Shinohara, and S. Iijima. Carbon. {\bf 44}, (2006), 1853. 
\bi{34} M. P. Siegal, D. L. Overmyer, and P. P. Provencio. Appl. Phys. Lett. {\bf 80}, (2002), 2171. 
\bi{35} C. L. Cheung, A. Kurtz, H. Park, and C. M. Lieber. J. Phys. Chem. B. {\bf 106}, (2002), 2429.
\bi{36} H. Kataura, Y. Kumazawa, Y. Maniwa, Y. Ohtsuka, R. Sen, S. Suzuki and Y. Achiba. Carbon. {\bf 38}, (2000), 1691. 
\bi{37} S. Bandow, S. Asaka, Y. Saito, A. M. Rao, L. Grigorian, E. Richter, and P. Eklund. Phys. Rev. Lett. {\bf 80}, (1998), 3779. 
\bi{38} H. Yasuoka,  R. Takahama,  M. Kaneda, and K. Suga.  Phys. Rev. E. {\bf 92}, (2015), 063001. 
\bi{39} H. Zhu, K. Suenaga, J. Wei, K. Wang, and D. Wu. J. Cryst. Growth. {\bf 310}, (2008), 5473. 
\bi{40} S. Reich, L. Li, and J. Robertson. Chem. Phys. Lett. {\bf 421}, (2006), 469. 
\bi{41} P. A. S. Autreto, S. B. Legoas, M. Z. S. Flores, and D. S. Galvao. J. Chem. Phys. {\bf 133}, (2010), 124513. 
\bi{42} K. Mizutani and  H. Kohno. Appl. Phys. Lett. {\bf 108}, (2016), 263112.
\bi{43} A. K. Abramyan, N. M. Bessonov, L. V. Mirantsev, A. A. Chevrychkina. Eur. Phys. J. B. {\bf 91}, (2018), 48. 
\bi{44}  L. V. Mirantsev, M. L. Lyra. Phys. Lett. A. {\bf 380}, (2016), 1318.
\bi{45} O. Teleman, B. Jonsson, S. Engstrom, Mol. Phys. {\bf 60}, (1987), 193.
\bi{46} J. H. Walther, R. Jaffe, T. Halicioglu, and P. Koumoutsakos, J. Phys. Chem. B {\bf 105},(2001), 9980. 
\bi{47} J. A. Barker, R. A. Fisher, R. D. Watts. Mol. Phys. {\bf 21}, (1971), 657.
\bi{48} M. P. Allen and D. J. Tildesly. \emph {Computer Simulations of Liquids}. Clarendon Press, Oxford, 1989.
\bi{49} H. J. C. Berendsen, J. P. M.  Postma, W. F. van Gunsteren, A. DiNola, and J. R. Haak, J. Chem. Phys. {\bf 81}, (1984), 3684. 
\bi{50} J. Tersoff. Phys. Rev. B {\bf 37}, (1988), 6991.
\bi{51} L. Wang and H. Hu. Proc. Math. Phys. Eng. Sci. {\bf 470}, (2014), 20140087.
\bi{52} I. E. Tamm. \emph {Fundamentals of the theory of eltctricity}. (Nauka, Moscow, 1989).
\bi{53} A. Keerthi, A. K. Geim, A. Janardanan, A. P. Rooney, A. Esfandiar, S. Hu, S. A. Dar, I. V. Grigorieva, S. J. Haigh, F. C. Wang, R. Radha.
Nature. {\bf 558}, (2018), 420.
\bi{54} L. D. Landau and E. M. Lifshitz. \emph {Fluid Mechanics}(Volume 6 of A Course of Theoretical Physics . (Pergamon Press, Oxford, 1959).


\end{thebibliography}
\end{document}